\documentclass[a4paper,11pt]{article}
\usepackage{jheppub} 
\usepackage{subcaption}

\usepackage{hyperref}
\usepackage{xstring}
\usepackage{url}
\usepackage{bm}
\newcommand{\BR}{\mathrm{BR}}
\newcommand{\GeV}{\mathrm{GeV}}
\newcommand{\TeV}{\mathrm{TeV}}

\newcommand{\MSSM}{\mathrm{MSSM}}

\newcommand{\MSUSY}{M_{\mathrm{SUSY}}}
\newcommand{\MGUT}{M_{\mathrm{GUT}}}

\title{\boldmath Phenomenological Study of Lepton Flavor Violation in $Z$-Boson Decays with the Constrained MSSM Extended by the Type-II Seesaw Model}

\author[a, b]{Vael Hajahmad}
\author[c]{Murhaf Alsayed Ali}
\affiliation[a]{ Erzincan Binali Yıldırım University,\\
	Faculty of Arts and Sciences, Physics Department, Erzincan, Türkiye}
\affiliation[b]{Al-Furat University,\\
	Faculty of Sciences, Physics Department,  Deir ez-Zor, Syria}
\affiliation[c]{Idlib University,\\
	Faculty of Sciences, Physics Department, Idlib, Syria}
\emailAdd{whahmad@erzincan.edu.tr}
\emailAdd{murhaf\_alsayed\_ali@idlib.edu.sy}

\abstract{This paper predicts lepton flavor violation (LFV) in $Z$-boson decays within the framework of the constrained minimal supersymmetric standard model (CMSSM) extended by the type-II seesaw model. The branching ratios of the decays $Z \to l_i l_j$ are predicted in this model, where $l_i$ and $l_j$ are charged leptons of different flavors. After fitting the experimental mass limits of neutrinos and supersymmetric particles; and imposing recent experimental constraints on $l_i \to l_j\gamma$ decays. We find that  the branching ratios of LFV $Z$-boson decays are of the order $2\times 10^{-12}$ for the $\mu\tau$ channel, $3 \times 10^{-12}$ for the $e \tau$ channel and $5 \times 10^{-18}$ for the $e\mu$ one. The theoretical predictions for the branching ratios are several orders of magnitude below the recent experimental limits, implying a very low probability of observing LFV $Z$-boson decays in future experiments.}

\providecommand{\keywords}[1]
{
	\small	
	\textbf{\textit{Keywords:}} #1
}
	\begin{document} 
		\maketitle
		\flushbottom
		
		\section{Introduction}

The Standard Model (SM) of elementary particle physics is the most successful theory for describing many physical phenomena over the range of energy scales currently accessible to experiments. However, the SM cannot be regarded as a complete theory because it does not generate neutrino masses \cite{ref1}. One of the most important open problems in particle physics is the origin of the tiny neutrino masses, which were discovered through neutrino-oscillation experiments \cite{ref2,ref3,ref4,ref5,ref6}. These experiments show that lepton flavor symmetry is not conserved in the neutrino sector \cite{ref7}. The upper limit on the neutrino mass from the Karlsruhe Tritium Neutrino Experiment (KATRIN) was estimated to be $[0.45-1.1]~\mathrm{eV}$ at $90\%$ CL \cite{refnutrino, ref8, ref9}.

Theoretically, implementing the seesaw mechanism in the SM and in supersymmetric (SUSY) models appears to be one of the simplest and most motivated solutions to the neutrino-mass problem \cite{ref10}. If neutrinos are Majorana particles, their mass at low energy is described by the unique dimension-5 operator \cite{ref11,ref12,ref13}
\begin{equation}
	\mathcal{L}_{m_\nu}=\frac{k}{\Lambda}LLHH ,
	\label{eq:weinberg}
\end{equation}
where $\Lambda$ is the high-energy scale of new physics, $k$ is a dimensionless coupling constant, $L$ is a lepton doublet, and $H$ is a Higgs doublet \cite{ref11}. For renormalizable interactions, three seesaw mechanisms can be realized at tree level. They are known as type-I, type-II, and type-III seesaw mechanisms and differ by the nature of their seesaw messengers. In the type-I seesaw, heavy fermionic singlets, often referred to as right-handed neutrinos, are exchanged \cite{ref14,ref15,ref16,ref17,ref18}. In the type-II seesaw, heavy $SU(2)_L$ scalar triplets with hypercharge two are exchanged \cite{ref18,ref19,ref20,ref21,ref22,ref23,ref24}. In the type-III seesaw, neutrino masses are generated through the exchange of $SU(2)_L$ fermionic triplets with zero hypercharge \cite{ref10,ref25,ref26,ref27}.

Supersymmetry at the TeV scale is one of the most promising candidates for new physics beyond the Standard Model (BSM). It can prevent the Higgs-boson mass from acquiring large quadratically divergent corrections, can enable successful gauge-coupling unification, and can provide a viable dark-matter candidate when exact $R$ parity is assumed, with the lightest neutralino as the dark-matter particle \cite{ref2}.

The seesaw mechanism is a useful SUSY extension for investigating light neutrino masses \cite{ref28}. It implies a new-physics energy scale of order $10^{14}~\GeV$ \cite{ref29}. Consequently, the supersymmetric seesaw mechanism may induce lepton-flavor-violating decays.

LFV decays in the SM with heavy neutrinos have extremely small rates and are therefore invisible experimentally. For example, the branching ratios for $Z\to e\tau$ and $Z\to e\mu$ are of order $10^{-54}$, while that for $Z\to \mu\tau$ is about $10^{-60}$. Charged LFV processes are forbidden in the SM with massless neutrinos \cite{ref7}. This makes LFV decays an attractive window for probing new-physics phenomena \cite{ref30}. Consequently, these processes can be studied both experimentally and theoretically within BSM theories. To date, no LFV signal has been observed experimentally; therefore, only upper bounds for various LFV processes have been reported \cite{ref31}.

The experimental upper limits on the branching ratios of the three $Z$-boson decay channels, $Z\to e\tau$, $Z\to\mu\tau$, and $Z\to e\mu$, are listed in table~\ref{tab:zlfv-limits}. These limits are based on recent results from the ATLAS experiment, which analyzed proton-proton collision data collected at a center-of-mass energy of $\sqrt{s}=13~\TeV$ and an integrated luminosity of $139~\mathrm{fb}^{-1}$ \cite{ref32,ref33}, as well as on data from the LEP collider.

At future $e^+e^-$ colliders such as CEPC and FCC-ee, and at the High-Luminosity Large Hadron Collider (HL-LHC), the expected sensitivities to the upper limits of the branching ratios for $Z\to e\tau$, $Z\to\mu\tau$, and $Z\to e\mu$ are also listed in table~\ref{tab:zlfv-limits}. These colliders are planned to operate near the $Z$ pole (``Tera-$Z$'' colliders). At these facilities, the number of collected $Z$-decay events is expected to be of order $10^{12}$, about a factor of $10^3$ larger than the LHC yield and $10^6$ larger than that at the LEP collider \cite{ref31,ref34,ref35}.

\begin{table}[htbp]
	\centering
	\caption{Experimental upper limits and expected sensitivities for the branching ratios of lepton-flavor-violating $Z$-boson decays.}
	\label{tab:zlfv-limits}
	\begin{tabular}{|l|c|c|c|c|}
		\hline
		Collider & LEP (95\% CL) & LHC (95\% CL) & HL-LHC & FCC-ee/CEPC \\
		\hline
		\hline
		$\BR(Z\to e\tau)$  & $9.8\times10^{-6}$ \cite{ref34} & $7.0\times10^{-6}$ \cite{ref32} & $10^{-6}$ \cite{ref35} & $10^{-9}$ \cite{ref31,ref34} \\
		\hline
		$\BR(Z\to\mu\tau)$ & $1.2\times10^{-5}$ \cite{ref34} & $7.20\times10^{-6}$ \cite{ref32} & $10^{-6}$ \cite{ref35} & $10^{-9}$ \cite{ref31,ref34} \\
		\hline
		$\BR(Z\to e\mu)$   & $1.7\times10^{-6}$ \cite{ref34} & $2.62\times10^{-7}$ \cite{ref33} & $10^{-7}$ \cite{ref35} & $10^{-8}$--$10^{-10}$ \cite{ref31,ref34} \\
		\hline
	\end{tabular}
\end{table}

In this study, we focus on $Z$-boson decays into two different lepton flavors, $Z\to l_i l_j$, where $l_i,l_j=e,\mu,\tau$. The work is performed within the constrained minimal supersymmetric standard model (CMSSM) extended by the type-II seesaw mechanism through the addition of a supersymmetric scalar triplet field.

In the SUSY seesaw framework, large slepton flavor mixings induce LFV interactions $\bar{l}_i l_j V$ with $V=\gamma,Z$. As a result, a correlation exists between the branching ratios of $Z\to l_i l_j$ decays and those of the radiative two-body decays $l_i\to l_j\gamma$ \cite{ref7,ref29,ref36}. We take into account the experimental bounds on both the masses of SUSY particles and the radiative decays $l_i\to l_j\gamma$ in order to constrain the model parameters and then evaluate the branching ratios of $Z\to l_i l_j$.
		\section{MSSM-seesaw type-II model}

In the type-II seesaw mechanism, the coupling between leptons and the Higgs field is mediated by the exchange of a scalar $SU(2)_L$ triplet $\widehat{T}$ \cite{ref37,ref38,ref39}. The hypercharge of the scalar triplet $\widehat{T}$ is taken to be 1 \cite{ref10}. To obtain a complete representation of $SU(5)$, $\widehat{T}$ is embedded in a 15-plet \cite{ref10,ref11,ref37,ref38,ref40,ref41}. After $SU(5)$ is broken, the 15-plet decomposes under the Standard Model gauge group $SU(3)_C\times SU(2)_L\times U(1)_Y$ as
\[
15=\widehat{S}+\widehat{T}+\widehat{Z},
\qquad
\widehat{S}\sim(6,1,-2/3),\quad
\widehat{T}\sim(1,3,1),\quad
\widehat{Z}\sim(3,2,1/6).
\]
Two 15-plets, namely $15$ and $\overline{15}$, are necessary to explain the light neutrino masses and to avoid chiral anomalies \cite{ref25,ref37,ref38,ref42,ref43,ref44,ref45}. The scalar triplet $\widehat{T}$ has the correct quantum numbers to generate the dimension-five operator \cite{ref11}. After $SU(5)$ breaking, the superpotential below the grand-unification scale $\MGUT$ is
\begin{align}
	W &= W_{\MSSM}+W_{\mathrm{SeesawII}},
	\label{eq:superpotential-total}\\
	W_{\mathrm{SeesawII}} &=
	\frac{1}{\sqrt{2}}\left(Y_T\widehat{L}\widehat{T}\widehat{L}
	+Y_S\widehat{d}\widehat{S}\widehat{d}\right)
	+Y_Z\widehat{d}\widehat{Z}\widehat{L}
	+\frac{1}{\sqrt{2}}\left(\lambda_1\widehat{H}_d\widehat{T}\widehat{H}_d
	+\lambda_2\widehat{H}_u\widehat{\overline{T}}\widehat{H}_u\right)
	\nonumber\\
	&\quad +M_T\widehat{T}\widehat{\overline{T}}
	+M_Z\widehat{Z}\widehat{\overline{Z}}
	+M_S\widehat{S}\widehat{\overline{S}} .
	\label{eq:superpotential-seesaw}
\end{align}
Here $\widehat{T}=(\widehat{T}^{0},\widehat{T}^{+},\widehat{T}^{++})$ and $\widehat{\overline{T}}=(\widehat{\overline{T}}^{0},\widehat{\overline{T}}^{-},\widehat{\overline{T}}^{--})$. The dimensionless, unflavored couplings $\lambda_1$ and $\lambda_2$ are the superpotential couplings of the triplets to the Higgs superfields. The parameter $M_T$ denotes the heavy-triplet mass scale, and $Y_T$ is the triplet Yukawa matrix \cite{ref10,ref11,ref42,ref46,ref47}. The superfields of particles in the MSSM-seesaw type-II model are shown in table~\ref{tab:superfields}.

\begin{table}[htbp]
	\centering
	\caption{Chiral superfields and their quantum numbers in the MSSM-seesaw type-II model. Here $\widehat{Q}$, $\widehat{L}$, $\widehat{H}_d$, and $\widehat{H}_u$ are the superfields of left-handed quarks, left-handed leptons, down-type Higgs, and up-type Higgs, respectively. The superfields $\widehat{d}$, $\widehat{u}$, and $\widehat{l}$ are the superfields of right-handed down-type quarks, right-handed up-type quarks, and right-handed leptons, respectively.}
	\label{tab:superfields}
	\begin{tabular}{|c|c|c|c|c|c|} 
		\hline 
		SF & Spin 0 & Spin \(\frac{1}{2}\) & Generations & \((U(1)\otimes\, \text{SU}(2)\otimes\, \text{SU}(3))\) \\ 
		\hline 
		\hline
		\(\hat{Q}\) & \(\tilde{q}\) & \(q\) & 3 & \((\frac{1}{6},{  2},{  3}) \) \\
		\hline 
		\(\hat{L}\) & \(\tilde{L}\) & \(L\) & 3 & \((-\frac{1}{2},{  2},{  1}) \) \\ 
		\hline
		\hline
		\(\hat{H}_d\) & \(H_d\) & \(\tilde{H}_d\) & 1 & \((-\frac{1}{2},{  2},{  1}) \) \\
		\hline 
		\(\hat{H}_u\) & \(H_u\) & \(\tilde{H}_u\) & 1 & \((\frac{1}{2},{  2},{  1}) \) \\ 
		\hline
		\hline
		\(\hat{d}\) & \(\tilde{d}_R^*\) & \(d_R^*\) & 3 & \((\frac{1}{3},{  1},{  \overline{3}}) \) \\ 
		\hline
		\(\hat{u}\) & \(\tilde{u}_R^*\) & \(u_R^*\) & 3 & \((-\frac{2}{3},{  1},{  \overline{3}}) \) \\
		\hline 
		\(\hat{l}\) & \(\tilde{l}_R^*\) & \(l_R^*\) & 3 & \((1,{  1},{  1}) \) \\ 
		\hline
		\hline
		\(\hat{T}\) & \(\tilde{T}\) & \(T\) & 1 & \((1,{  3},{  1}) \) \\ 
		\hline
		\(\hat{\bar{T}}\) & \(\tilde{\bar{T}}\) & \(\bar{T}\) & 1 & \((-1,{  3},{  1}) \) \\ 
		\hline
		\(\hat{S}\) & \(\tilde{S}\) & \(S\) & 1 & \((-\frac{2}{3},{  1},{  6}) \) \\ 
		\hline
		\(\hat{\bar{S}}\) & \(\tilde{\bar{S}}^*\) & \(\bar{S}^*\) & 1 & \((\frac{2}{3},{  1},{  \overline{6}}) \) \\ 
		\hline
		\(\hat{Z}\) & \(\tilde{Z}\) & \(Z\) & 1 & \((\frac{1}{6},{  2},{  3}) \) \\ 
		\hline
		\(\hat{\bar{Z}}\) & \(\tilde{\bar{Z}}\) & \(\bar{Z}\) & 1 & \((-\frac{1}{6},{  2},{  \overline{3}}) \) \\ 
		\hline 
	\end{tabular} 
\end{table}

At the heavy scalar-triplet scale $M_T$, below $\MGUT$, the Weinberg operator is given by \cite{ref48}
\begin{equation}
	\frac{1}{2}\kappa_\nu\,\widehat{L}\widehat{L}\widehat{H}_u\widehat{H}_u ,
	\label{eq:kappa-operator}
\end{equation}
where
\[
\kappa_\nu=\frac{\lambda_2}{M_T}Y_T .
\]

In the MSSM-seesaw type-II model, the soft supersymmetry-breaking terms for the heavy triplets are \cite{ref25,ref47,ref48}
\begin{align}
	-\mathcal{L}_{\mathrm{soft}} &=
	\mathcal{L}_{\mathrm{soft-MSSM}}
	+\frac{1}{\sqrt{2}}\left(A_TY_T\tilde{L}\tilde{T}\tilde{L}
	+A_SY_S\tilde{d}^{\,c}\tilde{S}\tilde{d}^{\,c}\right)
	+\frac{1}{\sqrt{2}}\left(A_1\lambda_1H_d\tilde{T}H_d
	+A_2\lambda_2H_u\tilde{\overline{T}}H_u\right)
	\nonumber\\
	&\quad +A_ZY_Z\tilde{d}^{\,c}\tilde{Z}\tilde{L}
	+B_TM_T\tilde{T}\tilde{\overline{T}}
	+B_ZM_Z\tilde{Z}\tilde{\overline{Z}}
	+B_SM_S\tilde{S}\tilde{\overline{S}}
	+\mathrm{h.c.}
	\nonumber\\
	&\quad +m_T^2\tilde{T}^\dagger\tilde{T}
	+m_{\overline{T}}^2\tilde{\overline{T}}^{\,\dagger}\tilde{\overline{T}}
	+m_S^2\tilde{S}^{\dagger}\tilde{S}
	+m_{\overline{S}}^2\tilde{\overline{S}}^{\,\dagger}\tilde{\overline{S}}
	+m_Z^2\tilde{Z}^{\dagger}\tilde{Z}
	+m_{\overline{Z}}^2\tilde{\overline{Z}}^{\,\dagger}\tilde{\overline{Z}} .
	\label{eq:soft-seesaw}
\end{align}
The parameters $m_T^2$, $m_S^2$, $m_Z^2$, $m_{\overline{T}}^2$, $m_{\overline{S}}^2$, and $m_{\overline{Z}}^2$ denote the soft SUSY-breaking mass-squared terms for the scalar triplets $(\tilde{T},\tilde{\overline{T}})$, scalar singlets $(\tilde{S},\tilde{\overline{S}})$, and scalar doublets $(\tilde{Z},\tilde{\overline{Z}})$, respectively. The parameters $A_T$, $A_S$, $A_Z$, $A_1$, and $A_2$ represent the trilinear couplings, while $B_T$, $B_Z$, and $B_S$ denote the bilinear couplings \cite{ref25}. The slepton and gaugino sectors of $\mathcal{L}_{\mathrm{soft,MSSM}}$ are
\begin{align}
	-\mathcal{L}_{\mathrm{soft,Slepton}} &=
	\sum_{i,j=\mathrm{gen}}(m_{\tilde{L}}^2)_{ij}\tilde{L}_i^{*}\tilde{L}_j
	+\sum_{i,j=\mathrm{gen}}(m_{\tilde{l}_R}^2)_{ij}\tilde{l}_{Ri}^{*}\tilde{l}_{Rj}
	+\sum_{i,j=\mathrm{gen}}(m_{\tilde{L}}^2)_{ij}\tilde{\nu}_{Li}^{*}\tilde{\nu}_{Lj}
	\nonumber\\
	&\quad
	+\sum_{i,j=\mathrm{gen}}A^l_{ij}Y^l_{ij}\tilde{l}_{Ri}^{*}\tilde{L}_jH_d+\mathrm{h.c.},
	\label{eq:soft-slepton}\\
	-\mathcal{L}_{\mathrm{soft,gaugino}} &=
	\frac{1}{2}M_3\tilde{G}\tilde{G}
	+\frac{1}{2}M_2\tilde{W}\tilde{W}
	+\frac{1}{2}M_1\tilde{B}\tilde{B}
	+\mathrm{h.c.}
	\label{eq:soft-gaugino}
\end{align}
Here $M_1$, $M_2$, and $M_3$ are the soft SUSY-breaking mass terms for the gaugino masses (bino, wino, and gluino). The matrices $m_{\tilde{L}}^2$ and $m_{\tilde{l}_R}^2$ are the soft SUSY-breaking mass-squared terms for supersymmetric leptons. The indices $i$ and $j$ denote generations. We define $T^l_{ij}=A^l_{ij}Y^l_{ij}$, where $A^l_{ij}$ are the trilinear coupling terms, and h.c. denotes Hermitian conjugation \cite{ref48,ref49,ref50}. The squark and Higgs terms are determined in the same way as in the MSSM \cite{ref50}.

In the MSSM-seesaw type-II model, after electroweak symmetry breaking (EWSB), the mass matrices for neutralinos, sleptons, and sneutrinos are introduced as follows, taking into account that the sneutrinos are separated into CP-odd and CP-even states:
\begin{equation}
	\tilde{\nu}_L=\frac{1}{\sqrt{2}}\left(\phi_L+i\sigma_L\right).
	\label{eq:sneutrino-decomposition}
\end{equation}
The CP-odd sneutrino mass matrix $m_{\tilde{\nu}^I}^2$ is
\begin{align}
	m_{\tilde{\nu}^I}^{2}=m_{\sigma_L\sigma_L}
	&=\frac{1}{8}\left(g_1^2+g_2^2\right)\left(-v_u^2+v_d^2\right)+\Re\left(m_{\tilde{L}}^2\right)
	\nonumber\\
	&\quad
	+\frac{1}{32}v_u^4\left[
	2\Re(\kappa_\nu\kappa_\nu^*)
	+2\Re(\kappa_\nu\kappa_\nu^\dagger)
	+2\Re(\kappa_\nu^t\kappa_\nu^*)
	+2\Re(\kappa_\nu^t\kappa_\nu^\dagger)\right]
	\nonumber\\
	&\quad
	+v_d\left[-4\left(4v_d\Re(Y_T\lambda_1^*)-v_u\mu^*
	(\kappa_\nu+\kappa_\nu^t+\kappa_{\nu,o1}+\kappa_{\nu,p1})\right)
	+8v_u\mu(\kappa_{\nu,o1}^*+\kappa_{\nu,p1}^*)\right].
	\label{eq:cpodd-sneutrino}
\end{align}
Here $g_1$ and $g_2$ are the gauge coupling constants of the electromagnetic $U(1)_Y$ and weak $SU(2)_L$ interactions, respectively. The parameters $v_u$ and $v_d$ denote the vacuum expectation values of $H_u$ and $H_d$, with $\tan\beta=v_u/v_d$. The indices $o1$ and $p1$ refer to the first generation. The matrix $m_{\tilde{\nu}^I}^2$ is diagonalized by $Z^I$:
\begin{equation}
	Z^I m_{\tilde{\nu}^I}^{2} Z^{I,\dagger}
	=\mathrm{diag}\left(m_{\tilde{\nu}_1^I}^2,m_{\tilde{\nu}_2^I}^2,m_{\tilde{\nu}_3^I}^2\right).
	\label{eq:cpodd-diag}
\end{equation}
The CP-even sneutrino mass matrix $m_{\tilde{\nu}^R}^2$ is
\begin{align}
	m_{\tilde{\nu}^R}^{2}=m_{\phi_L\phi_L}
	&=\frac{1}{8}\left(g_1^2+g_2^2\right)\left(-v_u^2+v_d^2\right)+\Re\left(m_{\tilde{L}}^2\right)
	\nonumber\\
	&\quad
	+\frac{1}{32}v_u^4\left[
	2\Re(\kappa_\nu\kappa_\nu^*)
	+2\Re(\kappa_\nu\kappa_\nu^\dagger)
	+2\Re(\kappa_\nu^t\kappa_\nu^*)
	+2\Re(\kappa_\nu^t\kappa_\nu^\dagger)\right]
	\nonumber\\
	&\quad
	+v_d\left[4\left(4v_d\Re(Y_T\lambda_1^*)-v_u\mu^*
	(\kappa_\nu+\kappa_\nu^t+\kappa_{\nu,o1}+\kappa_{\nu,p1})\right)
	-8v_u\mu(\kappa_{\nu,o1}^*+\kappa_{\nu,p1}^*)\right].
	\label{eq:cpeven-sneutrino}
\end{align}
The matrix $m_{\tilde{\nu}^R}^2$ is diagonalized by $Z^R$:
\begin{equation}
	Z^R m_{\tilde{\nu}^R}^{2}Z^{R,\dagger}
	=\mathrm{diag}\left(m_{\tilde{\nu}_1^R}^2,m_{\tilde{\nu}_2^R}^2,m_{\tilde{\nu}_3^R}^2\right).
	\label{eq:cpeven-diag}
\end{equation}
The terms in the second and third lines of eqs.~\eqref{eq:cpodd-sneutrino} and \eqref{eq:cpeven-sneutrino} are newly introduced by the MSSM-seesaw type-II model.

The charged-slepton mass-squared matrix is
\begin{equation}
	M_{\tilde{l}}^2=
	\begin{pmatrix}
		M_{LL} & M_{LR}\\
		M_{RL} & M_{RR}
	\end{pmatrix},
	\label{eq:slepton-mass-block}
\end{equation}
with
\begin{align}
	M_{LL} &=
	\frac{1}{2}v_d^2Y_l^\dagger Y_l
	+\frac{1}{8}(-g_2^2+g_1^2)(-v_u^2+v_d^2)
	+m_{\tilde{L}}^2,
	\label{eq:mll}\\
	M_{RR} &=
	\frac{1}{2}v_d^2Y_lY_l^\dagger
	+\frac{1}{4}g_1^2(-v_d^2+v_u^2)
	+m_{\tilde{l}_R}^2,
	\label{eq:mrr}\\
	M_{LR} &=
	\frac{1}{\sqrt{2}}\left(v_dT_l^\dagger-v_u\mu Y_l^\dagger\right),
	\label{eq:mlr}\\
	M_{RL} &=M_{LR}^\dagger=
	\frac{1}{\sqrt{2}}\left(v_dT_l-v_uY_l\mu^*\right).
	\label{eq:mrl}
\end{align}
Here $\mu$ is the Higgsino mass parameter. This matrix is diagonalized by $Z^E$:
\begin{equation}
	Z^E M_{\tilde{l}}^2 Z^{E,\dagger}
	=\mathrm{diag}\left(m_{\tilde{l}_1}^2,\ldots,m_{\tilde{l}_6}^2\right).
	\label{eq:slepton-diag}
\end{equation}
The sources of LFV are the off-diagonal entries of the $3\times3$ soft supersymmetry-breaking matrices $m_{\tilde{L}}^2$ and $m_{\tilde{l}_R}^2$.

The neutralino mass matrix is
\begin{equation}
	m_{\tilde{\chi}^{0}}=
	\begin{pmatrix}
		M_1 & 0 & -\frac{1}{2}g_1v_d & \frac{1}{2}g_1v_u\\
		0 & M_2 & \frac{1}{2}g_2v_d & -\frac{1}{2}g_2v_u\\
		-\frac{1}{2}g_1v_d & \frac{1}{2}g_2v_d & 0 & -\mu\\
		\frac{1}{2}g_1v_u & -\frac{1}{2}g_2v_u & -\mu & 0
	\end{pmatrix}.
	\label{eq:neutralino-mass}
\end{equation}
This matrix is diagonalized by $N$:
\begin{equation}
	N^*m_{\tilde{\chi}^{0}}N^\dagger
	=\mathrm{diag}\left(m_{\tilde{\chi}_1^0},m_{\tilde{\chi}_2^0},m_{\tilde{\chi}_3^0},m_{\tilde{\chi}_4^0}\right).
	\label{eq:neutralino-diag}
\end{equation}
The chargino mass matrix is
\begin{equation}
	m_{\tilde{\chi}^{\pm}}=
	\begin{pmatrix}
		M_2 & \frac{1}{\sqrt{2}}g_2v_u\\
		\frac{1}{\sqrt{2}}g_2v_d & \mu
	\end{pmatrix}.
	\label{eq:chargino-mass}
\end{equation}
This matrix is diagonalized by $U$ and $V$:
\begin{equation}
	U^*m_{\tilde{\chi}^{-}}V^\dagger
	=\mathrm{diag}\left(m_{\tilde{\chi}_1^\pm},m_{\tilde{\chi}_2^\pm}\right).
	\label{eq:chargino-diag}
\end{equation}

\begin{figure}[htbp]
	\centering
	\includegraphics[width=0.95\textwidth]{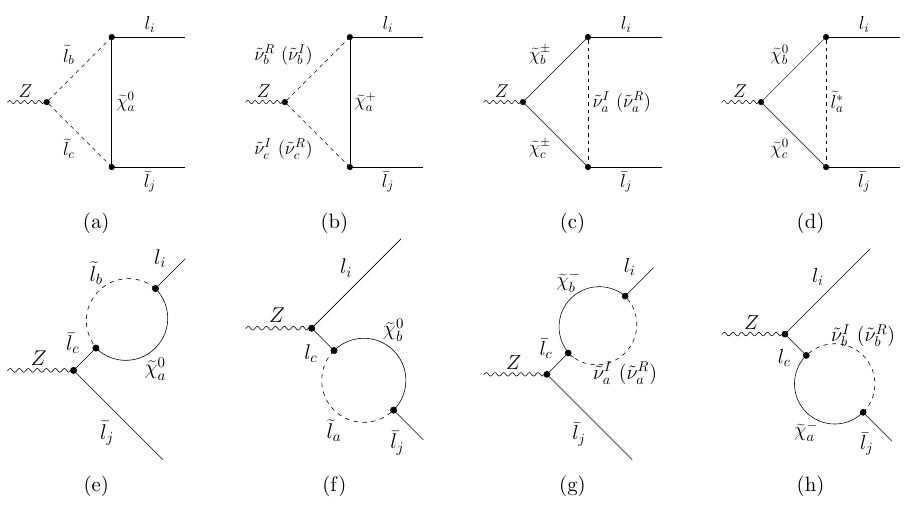}
	\caption{One-loop Feynman diagrams contributing to $\BR(Z\to l_i\bar{l}_j)$ in the MSSM-seesaw type-II model.}
	\label{fig:one-loop-zlfv}
\end{figure}

The charged sleptons and CP-odd (CP-even) sneutrinos exhibit flavor mixing. Charged-slepton mixing induces the flavor-changing neutral-current couplings $\tilde{\chi}^0 l \tilde{l}$ and $Z\tilde{l}\tilde{l}$, while CP-odd (CP-even) sneutrino mixing induces the flavor-changing charged-current coupling $\tilde{\chi}^+l\tilde{\nu}$. These flavor-changing couplings contribute to LFV $Z$ decays, $Z\to l_i l_j$ \cite{ref29,ref51}, as shown in figure~\ref{fig:one-loop-zlfv}.

The light-neutrino mass matrix is generated by electroweak symmetry breaking as \cite{ref37,ref38,ref52}
\begin{equation}
	m_\nu=\frac{1}{2}v_u^2\kappa_\nu .
	\label{eq:light-neutrino-kappa}
\end{equation}
Using eqs.~\eqref{eq:kappa-operator} and \eqref{eq:light-neutrino-kappa}, it can be written as
\begin{equation}
	m_\nu=\frac{v_u^2\lambda_2}{2M_T}Y_T .
	\label{eq:light-neutrino-yt}
\end{equation}
This matrix is diagonalized by $U^V$:
\begin{equation}
	U^{V,*}m_\nu U^{V,\dagger}
	=\mathrm{diag}\left(m_{\nu_1},m_{\nu_2},m_{\nu_3}\right).
	\label{eq:light-neutrino-diag}
\end{equation}
Equation~\eqref{eq:light-neutrino-yt} depends on the high-energy scale, $M_T\sim10^{14}~\GeV$. To calculate the neutrino mass at low energy, the values of $\lambda_2$, $M_T$, and $Y_T$ are needed as input parameters at the GUT scale, approximately $10^{16}~\GeV$. Furthermore, eq.~\eqref{eq:light-neutrino-yt} shows that the $Y_T$ matrix is similar to the $Y_\nu$ matrix because both matrices are diagonalized by the mixing matrix $U^V$. Thus, the structure of the $Y_T$ matrix can be like that of the $Y_\nu$ matrix \cite{ref11,ref37,ref38,ref48}. In principle, the flavored Yukawa couplings $Y_S$ and $Y_Z$ are not determined by low-energy neutrino data. However, they induce LFV slepton-mass terms in the same way as the $Y_T$ matrix \cite{ref10}.

In the general MSSM framework, the LFV off-diagonal entries in the slepton mass matrices involve additional free parameters arising from the mechanism of supersymmetry breaking. To relate LFV in the slepton sector to the LFV encoded in $Y_T$, a specific supersymmetry-breaking scheme must be assumed \cite{ref11}. Here, we consider the constrained MSSM at the GUT scale. The assumed universal conditions are
\[
M_1=M_2=M_3=m_{1/2},
\]
for the soft gaugino masses,
\[
m_{\tilde{l}_R}^2=m_{\tilde{L}}^2=m_0^2I,\qquad
m_T^2=m_S^2=m_Z^2=m_{\overline{T}}^2=m_{\overline{S}}^2=m_{\overline{Z}}^2
=m_{H_u}^2=m_{H_d}^2=m_0^2,
\]
for the scalar and Higgs soft masses, and
\[
A_l=A_1=A_2=A_0,\qquad
T^l_{ij}=A^l_{ij}Y^l_{ij},\quad
T_1=A_1\lambda_1,\quad
T_2=A_2\lambda_2,
\]
for the trilinear terms. Here $m_0$ is the universal scalar mass, $A_0$ is the universal mass parameter for the trilinear terms, and $m_{1/2}$ is the common gaugino mass. In addition, $\tan\beta=v_u/v_d$ is fixed at the electroweak scale \cite{ref11,ref48}.

We assume a specific scenario for the trilinear terms of $T$, $S$, and $Z$, namely $T_S=T_Z=0$ and $T_T=A_0Y_T$, with $Y_S=Y_Z=0$. For the mass terms, we assume $M_T=M_S=M_Z$. The one-loop contribution to supersymmetric lepton mixing is represented by the renormalization-group equations (RGEs) \cite{ref48}, where LFV occurs in the left-handed slepton sector:
\begin{align}
		\label{eq:rge-lfv}
	\Delta m_{\tilde{L}}^2
	&=-\frac{3}{8\pi^2}m_0^2
	\left(3+\frac{A_0^2}{m_0^2}\right)
	Y_T^\dagger Y_T\log\left(\frac{\MGUT}{M_T}\right),
	\nonumber\\
	\Delta T_l^2
	&=-\frac{9}{16\pi^2}A_0Y_lY_T^\dagger Y_T
	\log\left(\frac{\MGUT}{M_T}\right), \\
	\Delta m_{\tilde{l}_R}^2&=0. \nonumber
\end{align}
The presence of $Y_T$ in eq.~\eqref{eq:rge-lfv} induces LFV in the left-handed slepton mass matrix. We also note that the right-handed slepton terms do not receive any contribution in the leading-logarithmic approximation. Furthermore, the trilinear coupling terms are suppressed by the charged-lepton masses \cite{ref11,ref48}.

It is necessary to define the structure of the triplet Yukawa coupling matrix $Y_T$ because of its contribution to LFV, as shown in eq.~\eqref{eq:rge-lfv}. We assume that the entries of $Y_T$ are real, so that $Y_T^\dagger Y_T=Y_T^tY_T$, in order to avoid possible constraints from lepton electric dipole moments. Therefore, it is useful and constructive to consider a geometrical interpretation of the Yukawa coupling matrix, in which its elements are interpreted in flavor space as the components of three generic neutrino vectors ($\bm{n}_{\mu}$, $\bm{n}_{e}$, $\bm{n}_{\tau}$). A Yukawa coupling matrix in flavor space can be written as \cite{ref52,ref53}
\begin{equation}
\begin{aligned}
	Y_T = \left(\begin{matrix}y_{T11}&y_{T12}&y_{T13}\\y_{T21}&y_{T22}&y_{T23}\\y_{T31}&y_{T32}&y_{T33}\\\end{matrix}\right)\equiv f\left(\bm{n}_{e}\ \ \bm{n}_{\mu}\ \ \bm{n}_{\tau}\right) 
\end{aligned}
	\label{eq:yt-vectors}
\end{equation}
where $f$ denotes the strength of the neutrino Yukawa coupling. The vectors $\bm{n}_{\mu}$, $\bm{n}_{e}$, and $\bm{n}_{\tau}$ represent three generic neutrino vectors in flavor space. The term $Y_T^tY_T$ in the RGEs of eq.~\eqref{eq:rge-lfv} is related to LFV processes. It can be expressed as
\begin{equation}
	Y_T^t\ Y_T = f\left(\begin{matrix}\bm{n}_{e}\\\bm{n}_
	{\mu}\\\bm{n}_{\tau}\\\end{matrix}\right)f
\left(\bm{n}_{e}\ \bm{n}_{\mu}\ \bm{n}_{\tau}\right){
	= f}^2\left(\begin{matrix}\left|\bm{n}_e\right|^2&\bm{n}_{e}
	\ .\ \bm{n}_{\mu}&\bm{n}_{e}\ .\ 
	\bm{n}_{\tau}\\{\bm{n}_{\mu}\ . \
		\bm{n}}_{e}\ &{|\bm{n}_\mu|}^2&
	\bm{n}_{\mu}\ .\ \bm{n}_{\tau}\\{
		\bm{n}_{\tau}\ .\ \bm{n}}_{e}\ &
	\bm{n}_{\tau}\ .\ \bm{n}_{\mu}\ &\left|\bm{n}_
	\tau\right|^2\\\end{matrix}\right)
	\label{eq:yt-product}
\end{equation}
Here $\bm{n}_{i} \cdot \bm{n}_{j}$  represents the dot product between the neutrino flavor vectors, $\bm{n}_{i} \cdot \bm{n}_{j}=|n_i||n_j|C_{ij}$, where $C_{ij}\equiv\cos\theta_{ij}$ is the cosine of the three neutrino flavor angles $C_{\tau\mu}$, $C_{e\mu}$, and $C_{\tau e}$. The angles are named according to the fact that $\cos\theta_{ij}$ controls LFV transitions in the $l_i-l_j$ sector. Therefore, the nine input parameters determining the $Y_T$ matrix can be taken as three moduli of the neutrino vectors $(|\bm{n}_e|,|\bm{n}_\mu|,|\bm{n}_\tau|)$, three relative flavor angles between these vectors $(\theta_{\mu e},\theta_{\tau e},\theta_{\tau\mu})$, and three additional angles $(\theta_1,\theta_2,\theta_3)$ determining the global rotation $O$ of the neutrino vectors without changing their relative angles. The matrix $O$ is an orthogonal rotation matrix and does not enter the product $Y_T^tY_T$ because $O^tO=I$; therefore, it does not affect the study of LFV \cite{ref52,ref53}. For real entries of $Y_T$, the Yukawa matrix can be written as the product
\begin{equation}
	Y_T=OA.
	\label{eq:yt-oa}
\end{equation}

The elements of $A$ are determined according to three possible scenarios: electron-tau, tau-muon, and electron-muon. In the electron-tau (ET) scenario, we substitute $C_{e\mu}=C_{\tau\mu}=0$ in eq.~\eqref{eq:yt-product}, implying that the vector pairs $(\bm{n}_\mu,\bm{n}_e)$ and $(\bm{n}_\mu,\bm{n}_\tau)$ are orthogonal. Thus,
\begin{equation}
	Y_T^tY_T=f^2
	\begin{pmatrix}
		|\bm{n}_e|^2 & 0 & \bm{n}_e\cdot \bm{n}_\tau\\
		0 & |\bm{n}_\mu|^2 & 0\\
		\bm{n}_\tau\cdot \bm{n}_e & 0 & |\bm{n}_\tau|^2
	\end{pmatrix}.
	\label{eq:yt-product-et}
\end{equation}
The Yukawa coupling matrix for the electron-tau scenario can be written as
\begin{equation}
	Y_T^{e\tau}=OA_{e\tau}
	=O\,f
	\begin{pmatrix}
		|\bm{n}_e| & 0 & |\bm{n}_\tau|C_{e\tau}\\
		0 & |\bm{n}_\mu| & 0\\
		0 & 0 & |\bm{n}_\tau|\sqrt{1-C_{e\tau}^{2}}
	\end{pmatrix}.
	\label{eq:yt-et}
\end{equation}
By calculating the product $Y_T^tY_T$, we recover eq.~\eqref{eq:yt-product-et}. In this case, the matrix $A_{e\tau}$ is
\begin{equation}
	A_{e\tau}=f
	\begin{pmatrix}
		|\bm{n}_e| & 0 & |\bm{n}_\tau|C_{e\tau}\\
		0 & |\bm{n}_\mu| & 0\\
		0 & 0 & |\bm{n}_\tau|\sqrt{1-C_{e\tau}^{2}}
	\end{pmatrix}.
	\label{eq:a-et}
\end{equation}
For the tau-muon (TM) scenario, we substitute $C_{e\mu}=C_{e\tau}=0$ in eq.~\eqref{eq:yt-product}, so that $(\bm{n}_\mu,\bm{n}_e)$ and $(\bm{n}_e,\bm{n}_\tau)$ are orthogonal. The resulting Yukawa coupling matrix is
\begin{equation}
	Y_T^{\tau\mu}=OA_{\tau\mu}
	=O\,f
	\begin{pmatrix}
		|\bm{n}_e| & 0 & 0\\
		0 & |\bm{n}_\mu| & |\bm{n}_\tau|C_{\tau\mu}\\
		0 & 0 & |\bm{n}_\tau|\sqrt{1-C_{\tau\mu}^{2}}
	\end{pmatrix}.
	\label{eq:yt-tm}
\end{equation}
In this scenario, the matrix $A_{\tau\mu}$ is
\begin{equation}
	A_{\tau\mu}=f
	\begin{pmatrix}
		|\bm{n}_e| & 0 & 0\\
		0 & |\bm{n}_\mu| & |\bm{n}_\tau|C_{\tau\mu}\\
		0 & 0 & |\bm{n}_\tau|\sqrt{1-C_{\tau\mu}^{2}}
	\end{pmatrix}.
	\label{eq:a-tm}
\end{equation}
For the electron-muon (EM) scenario, we substitute $C_{\tau\mu}=C_{e\tau}=0$ in eq.~\eqref{eq:yt-product}, so that $(\bm{n}_\mu,\bm{n}_\tau)$ and $(\bm{n}_\mu,\bm{n}_e)$ are orthogonal. The Yukawa coupling matrix becomes
\begin{equation}
	Y_T^{e\mu}=OA_{e\mu}
	=O\,f
	\begin{pmatrix}
		|\bm{n}_e|\sqrt{1-C_{e\mu}^{2}} & 0 & 0\\
		|\bm{n}_e|C_{e\mu} & |\bm{n}_\mu| & 0\\
		0 & 0 & |\bm{n}_\tau|
	\end{pmatrix}.
	\label{eq:yt-em}
\end{equation}
Thus,
\begin{equation}
	A_{e\mu}=f
	\begin{pmatrix}
		|\bm{n}_e|\sqrt{1-C_{e\mu}^{2}} & 0 & 0\\
		|\bm{n}_e|C_{e\mu} & |\bm{n}_\mu| & 0\\
		0 & 0 & |\bm{n}_\tau|
	\end{pmatrix}.
	\label{eq:a-em}
\end{equation}
The TM scenario, with $C_{e\tau}=C_{e\mu}=0$, can produce large rates for $\tau-\mu$ transitions but always gives negligible contributions to LFV in the $e\mu$ and $e\tau$ sectors. The ET scenario, with $C_{\tau\mu}=C_{e\mu}=0$, can yield sizable rates for $\tau-e$ transitions but gives negligible contributions to the $e\mu$ and $\tau\mu$ channels. The EM scenario, with $C_{\tau\mu}=C_{e\tau}=0$, can produce large rates only for the $\mu-e$ transitions \cite{ref53}.

	    \section{The LFV decays of the $Z$ boson ($Z\to l_i l_j$)}

The Lagrangian for LFV in $Z$-boson decays can be written as \cite{ref45,ref54}
\begin{equation}
	\mathcal{L}_{Zl_il_j}=
	\bar{l}_j\left[\gamma^\mu(A_1^L P_L+A_1^R P_R)
	+p^\mu(A_2^L P_L+A_2^R P_R)\right]l_i Z_\mu ,
	\label{eq:zlfv-lagrangian}
\end{equation}
where $P_{L,R}=\frac{1}{2}(1\pm\gamma^5)$ are chirality projectors, $l_i$ and $l_j$ represent lepton flavors, and $p$ is the four-momentum of $l_j$. The coefficients $A_1^L$, $A_1^R$, $A_2^L$, and $A_2^R$ can be obtained from the amplitudes of the Feynman diagrams shown in figure~\ref{fig:one-loop-zlfv}. By neglecting charged-lepton masses, the branching ratio for $Z$-boson decays is \cite{ref45,ref54,ref55}
\begin{equation}
	\BR(Z\to l_i l_j)
	=\BR(Z\to l_i\bar{l}_j)+\BR(Z\to \bar{l}_i l_j)
	=\frac{\Gamma(Z\to l_i\bar{l}_j)+\Gamma(Z\to \bar{l}_i l_j)}{\Gamma_Z}.
	\label{eq:br-sum}
\end{equation}
Here $\Gamma_Z$ is the total decay width of the $Z$ boson, $\Gamma_Z=2.4952~\GeV$ \cite{ref56}, while the decay width is \cite{ref45,ref55}
\begin{equation}
	\Gamma(Z\to l_i l_j)
	=\frac{m_Z}{48\pi}
	\left[
	2\left(|A_1^L|^2+|A_1^R|^2\right)
	+\frac{m_Z^2}{4}\left(|A_2^L|^2+|A_2^R|^2\right)
	\right].
	\label{eq:width-zlfv}
\end{equation}
The final expression for the branching ratio is therefore
\begin{equation}
	\BR(Z\to l_i l_j)
	=\frac{m_Z}{48\pi\Gamma_Z}
	\left[
	2\left(|A_1^L|^2+|A_1^R|^2\right)
	+\frac{m_Z^2}{4}\left(|A_2^L|^2+|A_2^R|^2\right)
	\right].
	\label{eq:br-zlfv}
\end{equation}
The coefficients $A_1^{L/R}$ and $A_2^{L/R}$ are sums of the corresponding coefficients from each Feynman diagram in figure~\ref{fig:one-loop-zlfv}:
\[
A_1^{L/R}=A_{1a}^{L/R}+A_{1b}^{L/R}+A_{1c}^{L/R}+A_{1d}^{L/R}
+A_{1e}^{L/R}+A_{1f}^{L/R}+A_{1g}^{L/R}+A_{1h}^{L/R},
\]
\[
A_2^{L/R}=A_{2a}^{L/R}+A_{2b}^{L/R}+A_{2c}^{L/R}+A_{2d}^{L/R}
+A_{2e}^{L/R}+A_{2f}^{L/R}+A_{2g}^{L/R}+A_{2h}^{L/R}.
\]
The neutralino-slepton loop contributions are derived from figure~\ref{fig:one-loop-zlfv}(a,d,e,f), while the chargino-sneutrino loop contributions are derived from figure~\ref{fig:one-loop-zlfv}(b,c,g,h).

The Feynman-diagram contributions from figures~\ref{fig:one-loop-zlfv}(a,b) are
\begin{align}
	A_1^{L(a,b)} &=2V_1^L V_2^R V_Z C_{00},
	\label{eq:a1-ab}\\
	A_2^{L(a,b)} &=2V_1^L V_2^L V_Z(C_0+C_1+C_2)M,
	\label{eq:a2-ab}\\
	A_1^R&=A_1^L(L\leftrightarrow R),\qquad
	A_2^R=A_2^L(L\leftrightarrow R).
	\label{eq:ar-ab}
\end{align}
For figure~\ref{fig:one-loop-zlfv}(a), the couplings are defined as
\[
V_1^L=\Gamma_{\tilde{\chi}_a^0 l_i \tilde{l}_b^{*}}^{L}\quad, \quad V_2^{L/R}=\Gamma_{\bar{l}_j \tilde{\chi}_a^{0} \tilde{l}_c}^{L/R} \quad,\quad
V_Z=\Gamma_{\tilde{l}_c \tilde{l}_b^{*}Z}
\]
Here $\Gamma_{\tilde{\chi}_a^0 l_i \tilde{l}_b^{*}}^{L}$ is the left-handed coupling of the neutralino-lepton-slepton vertex $(\tilde{\chi}^{0}-l-\tilde{l}^{*})$, $\Gamma_{\bar{l}_j \tilde{\chi}_a^{0} \tilde{l}_c}^{L/R}$ is the left(right)-handed coupling of the antilepton-neutralino-slepton vertex $(\tilde{\chi}^{0}-l-\tilde{l}^{*})$, and $\Gamma_{\tilde{l}_c \tilde{l}_b^{*}Z}$ is the coupling of the slepton-slepton-$Z$ vertex $(\tilde{l}-\tilde{l}^{*}-Z)$. The explicit forms of these couplings are given in appendix~\ref{sec:appendix}. In eq.~\eqref{eq:a2-ab}, $M$ represents the neutralino mass, $M=m_{\tilde{\chi}_a^0}$.

For figure~\ref{fig:one-loop-zlfv}(b), the couplings are
\[
V_1^L=\Gamma_{\tilde{\chi}_a^{+} l_i \tilde{\nu}_b^{I/R}}^{L} \quad,\quad
V_2^{L/R}=\Gamma_{l_j \tilde{\chi}_a^{-}\tilde{\nu}_c^{I/R}}^{L/R} \quad,\quad
V_Z=\Gamma_{\tilde{\nu}_b^{I}\tilde{\nu}_c^{R} Z}=-\Gamma_{\tilde{\nu}_c^{I}\tilde{\nu}_b^{R}Z}
\]
Here $\Gamma_{\tilde{\chi}_a^{+} l_i \tilde{\nu}_b^{I/R}}^{L}$ is the left-handed coupling of the chargino-lepton-(CP-odd/CP-even) sneutrino vertex, $\Gamma_{l_j \tilde{\chi}_a^{-}\tilde{\nu}_c^{I/R}}^{L/R}$ represents the left(right)-handed coupling of the lepton-chargino-(CP-odd/CP-even) sneutrino vertex, and $\Gamma_{\tilde{\nu}_c^{I}\tilde{\nu}_b^{R}Z}$ is the coupling of the CP-odd sneutrino--CP-even sneutrino--$Z$ vertex. In eq.~\eqref{eq:a2-ab}, $M$ now represents the chargino mass, $M=m_{\tilde{\chi}_a^\pm}$. The explicit forms of these couplings are given in appendix~\ref{sec:appendix}.

For figures~\ref{fig:one-loop-zlfv}(a,b), the parameters $C_0$, $C_{00}$, $C_1$, and $C_2$ denote the standard three-point functions as defined in the LoopTools package \cite{ref57,ref58}; they can also be calculated using the Mathematica package Package-X \cite{ref59}. The arguments of the $C$ functions for figures~\ref{fig:one-loop-zlfv}(a,b) are
\[
(0,m_Z^2,0,m_{\tilde{\chi}_a^0}^2,m_{\tilde{l}_c}^2,m_{\tilde{l}_b}^2)
\]
and
\[
(0,m_Z^2,0,m_{\tilde{\chi}_a^-}^2,m_{\tilde{\nu}_c^{R/I}}^2,m_{\tilde{\nu}_b^{I/R}}^2),
\]
respectively, with the external fermion masses set to zero.

The contributions from figures~\ref{fig:one-loop-zlfv}(c,d) are
\begin{align}
	A_1^{L(c,d)}&=V_1^L V_2^R
	\left[V_Z^L C_0 m_1m_2 - V_Z^R(B_0-2C_{00}+C_0)m_3^2\right],
	\label{eq:a1-cd}\\
	A_2^{L(c,d)}&=2V_1^L V_2^L
	\left[-V_Z^L C_1m_1+V_Z^R(C_0+C_1+C_2)m_2\right],
	\label{eq:a2-cd}\\
	A_1^R&=A_1^L(L\leftrightarrow R),\qquad
	A_2^R=A_2^L(L\leftrightarrow R).
	\label{eq:ar-cd}
\end{align}
For figure~\ref{fig:one-loop-zlfv}(c), the couplings are
\[
V_1^{L/R}=\Gamma_{\tilde{\chi}_b^{+} l_i \tilde{\nu}_a^{I/R}}^{L/R} \quad,\quad
V_2^{R/L}=\Gamma_{\bar{l}_j\tilde{\chi}_c^{-}\tilde{\nu}_a^{I/R}}^{R/L} \quad,\quad
V_Z^{L/R}=\Gamma_{\tilde{\chi}_c^{+}\tilde{\chi}_b^{-}Z}^{L/R}
\]
For figure~\ref{fig:one-loop-zlfv}(d), the couplings are
\[
V_1^L=\Gamma_{\tilde{\chi}_b^{0}l_i \tilde{l}_a^{*}}^{L} \quad,\quad
V_2^{R/L}=\Gamma_{\bar{l}_j\tilde{\chi}_c^{0}\tilde{l}_a}^{R/L} \quad,\quad
V_Z^{L/R}=\Gamma_{\tilde{\chi}_c^{0}\tilde{\chi}_b^{0}Z}^{L/R}
\]
Here $\Gamma_{\tilde{\chi}_c^{+}\tilde{\chi}_b^{-}Z}^{L/R}$ represents the chargino-chargino-$Z$ coupling, and $\Gamma_{\tilde{\chi}_c^{0}\tilde{\chi}_b^{0}Z}^{L/R}$ denotes the neutralino-neutralino-$Z$ coupling. These couplings are given in appendix~\ref{sec:appendix}.

For figure~\ref{fig:one-loop-zlfv}(c), the parameters $m_1$, $m_2$, and $m_3$ represent the masses of the chargino and CP-odd/CP-even sneutrino:
\[
m_1=m_{\tilde{\chi}_b^-},\qquad
m_2=m_{\tilde{\chi}_c^-},\qquad
m_3=m_{\tilde{\nu}_a^{I/R}}.
\]
For figure~\ref{fig:one-loop-zlfv}(d), $m_1$ and $m_2$ represent neutralino masses and $m_3$ is the slepton mass:
\[
m_1=m_{\tilde{\chi}_b^0},\qquad
m_2=m_{\tilde{\chi}_c^0},\qquad
m_3=m_{\tilde{l}_a}.
\]
The arguments of the $C$ functions for figures~\ref{fig:one-loop-zlfv}(c,d) are
\[
(m_Z^2,0,0,m_{\tilde{\chi}_c^-}^2,m_{\tilde{\chi}_b^-}^2,m_{\tilde{\nu}_a^{I/R}}^2)
\quad\mathrm{and}\quad
(m_Z^2,0,0,m_{\tilde{\chi}_c^0}^2,m_{\tilde{\chi}_b^0}^2,m_{\tilde{l}_a}^2),
\]
respectively, with the external fermion masses set to zero. The function $B_0$ denotes a two-point function. The arguments of the $B_0$ function for figures~\ref{fig:one-loop-zlfv}(c,d) are
\[
(m_Z^2,m_{\tilde{\chi}_b^-}^2,m_{\tilde{\chi}_c^-}^2)
\quad\mathrm{and}\quad
(m_Z^2,m_{\tilde{\chi}_b^0}^2,m_{\tilde{\chi}_c^0}^2),
\]
respectively.

The contributions of the Feynman diagrams in figures~\ref{fig:one-loop-zlfv}(e,f,g,h), where the $Z$ boson does not couple to a scalar particle and therefore $A_2^{L/R}=0$, are
\begin{align}
	A_1^L &=
	\frac{V_Z^L}{m_1^2-m_2^2}
	\left[
	- V_1^L V_2^R B_1m_2^2
	+ V_1^R V_2^R B_0m_1m_3
	- V_1^R V_2^L B_1m_1m_2
	+ V_1^L V_2^L B_0m_3m_2
	\right],
	\label{eq:a1-efgh}\\
	A_1^R&=A_1^L(L\leftrightarrow R).
	\label{eq:ar-efgh}
\end{align}
For figures~\ref{fig:one-loop-zlfv}(e,f), the couplings are
\[
V_1^{L/R}=\Gamma_{\tilde{\chi}_a^{0}l_i \tilde{l}_b^{*}}^{L/R}
=\Gamma_{\tilde{\chi}_b^{0}l_c \tilde{l}_a^{*}}^{L/R},
\]
\[
V_2^{R/L}=\Gamma_{\bar{l}_a \tilde{\chi}_i^{0}\tilde{l}_b}^{R/L}
=\Gamma_{\bar{l}\tilde{\chi}^{0}\tilde{l},jba}^{R/L} \quad,
\quad
V_Z^L=\Gamma_{\bar{l}_j l_c Z}^{L}
=\Gamma_{\bar{l}_c l_i Z}^{L}.
\]
Here $\Gamma_{\bar{l}lZ}^{L}$ is the left-handed antilepton-lepton-$Z$ coupling. Its explicit form is given in appendix~\ref{sec:appendix}. For figure~\ref{fig:one-loop-zlfv}(e), $m_1$, $m_2$, and $m_3$ denote lepton and neutralino masses:
\[
m_1=m_{l_i},\qquad m_2=m_{l_c},\qquad m_3=m_{\tilde{\chi}_a^0}.
\]
For figure~\ref{fig:one-loop-zlfv}(f), the masses are
\[
m_1=m_{l_j},\qquad m_2=m_{l_c},\qquad m_3=m_{\tilde{\chi}_b^0}.
\]

For figures~\ref{fig:one-loop-zlfv}(g,h), the couplings are
\[
V_1^{L/R}=\Gamma_{\tilde{\chi}_b^{+} l_i \tilde{\nu}_a^{I/R}}^{L/R}
=\Gamma_{\tilde{\chi}_a^{+} l_c \tilde{\nu}_b^{I/R}}^{L/R},
\]
\[
V_2^{R/L}=\Gamma_{\bar{l}_c\tilde{\chi}_b^{-}\tilde{\nu}_a^{I/R}}^{R/L}
=\Gamma_{\bar{l}_j\tilde{\chi}_a^{-}\tilde{\nu}_b^{I/R}}^{R/L} \quad,
\quad
V_Z^L=\Gamma_{\bar{l}lZ,jc}^{L}
=\Gamma_{\bar{l}_c l_i Z}^{L}.
\]
For figures~\ref{fig:one-loop-zlfv}(g,h), $m_1$, $m_2$, and $m_3$ represent the masses of the lepton and CP-odd/CP-even sneutrino and are defined as
\[
(m_1,m_2,m_3)=(m_{l_i},m_{l_c},m_{\tilde{\chi}_b^-})
\quad\mathrm{and}\quad
(m_1,m_2,m_3)=(m_{l_j},m_{l_c},m_{\tilde{\chi}_a^-}),
\]
respectively. The arguments of the $B$ functions for figures~\ref{fig:one-loop-zlfv}(e,f,g,h) are
\[
(m_{l_i}^2,m_{\tilde{\chi}_a^0}^2,m_{\tilde{l}_b}^2),\qquad
(m_{l_j}^2,m_{\tilde{\chi}_b^0}^2,m_{\tilde{l}_a}^2),
\]
\[
(m_{l_i}^2,m_{\tilde{\chi}_b^-}^2,m_{\tilde{\nu}_a^{I/R}}^2),\qquad
(m_{l_j}^2,m_{\tilde{\chi}_a^-}^2,m_{\tilde{\nu}_b^{I/R}}^2),
\]
respectively.
		\section{Numerical results and discussion}

In this section, the numerical results are obtained using the SARAH, SPheno, and FlavorKit packages with full two-loop RGEs \cite{ref43,ref44,ref60,ref61}. SARAH is a Mathematica package for building and analyzing SUSY and non-SUSY models; it creates source code for the SPheno tool. SPheno (Supersymmetric Phenomenology) is written in Fortran 90 and calculates the SUSY spectrum using low-energy data together with a user-supplied high-scale model as input, such as the MSSM, type-I seesaw, type-II seesaw, type-III seesaw, next-to-minimal MSSM, and other models. In addition, FlavorKit, available within SARAH/SPheno, can compute a wide range of flavor observables, including LFV in $Z$-boson, Higgs-boson, and tau-lepton decays \cite{ref45}.

According to the MSSM-seesaw type-II model, the final parameters in this study are
\[
Y_T,\; M_T,\; f,\; |\bm{n}_e|,\; |\bm{n}_\mu|,\; |\bm{n}_\tau|,\; C_{\tau\mu},\; C_{e\tau},\; C_{e\mu},\; m_{1/2},\; m_0,\; A_0,\; \tan\beta,\; \mathrm{sign}(\mu),\; \lambda_1,\; \lambda_2 .
\]
In our calculations, the soft-symmetry-breaking terms are constrained by several theoretical and experimental conditions. For example, the lightest supersymmetric particle of the model is the neutralino, and $R$ parity is conserved. In addition, the supersymmetric particle masses (charginos, sleptons, sneutrinos, and neutralinos), calculated with SPheno, must be above the recent experimental mass limits \cite{ref56}, as shown in table~\ref{tab:sparticle-mass}.

\begin{table}[htbp]
	\centering
	\caption{Experimental mass limits of supersymmetric particles \cite{ref56}.}
	\label{tab:sparticle-mass}
	\begin{tabular}{|l|c|}
		\hline
		Sparticle &Mass limits (GeV) \\
		\hline \hline
		Sleptons& $\ge 107$\\
		\hline
		Sneutrinos&$\ge 94$\\
		\hline
		Neutralinos&$\ge 46$\\
		\hline
		Charginos&$\ge 94$ \\
		\hline
	\end{tabular}
\end{table}

The supersymmetric-particle masses are related to both $m_{1/2}$ and $m_0$. Therefore, we first determine their minimum allowed values according to the previous conditions, obtaining $m_{1/2}=200~\GeV$ and $m_0=550~\GeV$. The ranges of input parameters considered in this study are shown in table~\ref{tab:input-ranges}.

\begin{table}[htbp]
	\centering
	\caption{Ranges of input parameters considered in this study.}
	\label{tab:input-ranges}
	\begin{tabular}{|l|c|}
		\hline
		Parameter &Values \\
		\hline \hline
		$m_0$ (GeV)& [550,1550]\\
		\hline
		$m_{1/2}$ (GeV)& [200,1000]\\
		\hline
		$cos(\theta_{ij})$& [0.087,0.87]\\
		\hline 
		$A_0$ (GeV)& 0\\
		\hline \hline
		$tan\beta$& 5\\
		\hline
		$\lambda_1=\lambda_2$ & 0.5\\
		\hline 
		 $\mathrm{sign}(\mu)$ & $>$ 0\\
		\hline
		$M_T~(\GeV)$& $5\times10^{13}$\\
		\hline
	\end{tabular}
\end{table}

From eqs.~\eqref{eq:light-neutrino-yt}, \eqref{eq:yt-et}, \eqref{eq:yt-tm}, and \eqref{eq:yt-em}, we estimate both the triplet Yukawa matrix elements and the scalar-triplet mass $M_T$ by imposing the light-neutrino mass limit, which is $<0.45~\mathrm{eV}$ at low energy \cite{refnutrino}. We can estimate the Yukawa matrix elements for any fixed neutrino mass as a function of the triplet mass for fixed Yukawa couplings. This estimate does not directly provide the correct Yukawa couplings because the neutrino masses are measured at low energy, whereas calculation of $m_\nu$ from eq.~\eqref{eq:light-neutrino-yt} requires $M_T$ and $Y_T$ as high-scale input parameters. Nevertheless, this estimate can be used to run the RGEs numerically and obtain the exact light-neutrino masses at low energy for these inputs. The difference between the obtained results and the desired values can be minimized numerically in a simple iterative procedure until convergence is achieved. This convergence can be reached in several steps as long as the Yukawa matrix elements satisfy $|(Y_T)_{ij}|<1$ \cite{ref41}.

In this work, sizable values of the triplet Yukawa couplings are considered, while they are still required to remain within the perturbative regime. Thus, the constraint on the maximum allowed entries of the Yukawa matrix is chosen as $|(Y_T)_{ij}|^2<4\pi$ \cite{ref52}; consequently, we obtain $M_T\geq 5\times10^{13}~\GeV$ and $\cos(\theta_{ij})\leq0.87$. We fix $\lambda_1=\lambda_2=0.5$ and $\mathrm{sign}(\mu)>0$, as in ref.~\cite{ref10}, for all numerical calculations. From eqs.~\eqref{eq:yt-et}, \eqref{eq:yt-tm}, and \eqref{eq:yt-em}, we have three Yukawa-matrix scenarios for LFV $Z$-boson decays. The values of $|\bm{n}_e|$, $|\bm{n}_\mu|$, and $|\bm{n}_\tau|$ for these scenarios are given in table~\ref{tab:yukawa-scenarios}. The GUT scale is fixed to $\MGUT=2.00\times10^{16}~\GeV$, while the supersymmetry-breaking scale is fixed to $\MSUSY=10^3~\GeV$ \cite{ref38}. We study $\BR(Z\to l_i l_j)$ with applying the constraints from the non-observation of ($l_i\to l_j\gamma$) decays.

\begin{table}[htbp]
	\centering
	\caption{Yukawa-matrix scenarios for the numerical calculation of $\BR(Z\to l_i l_j)$.}
	\label{tab:yukawa-scenarios}
	\begin{tabular}{|l|l|l|l|l|l|l|c|}
		\hline
		Scenario & $C_{\tau \mu}$ & $C_{e \tau}$ & $C_{e \mu}$ &$\left|\bm{n}_e\right|$ &$\left|\bm{n}_\mu \right|$ & $\left|\bm{n}_\tau \right|$ &\ Example \\
		\hline \hline
		MT &$C_{\tau \mu}$&\ 0 &\ 0 &\ 0.1 & \ 1 &\ 1 & $Y_T = f\left(\begin{matrix}0.1&0&0\\ 0&1&C_{\tau \mu}\\0&0&\sqrt{1-C_{\tau\mu}^2}\\\end{matrix}\right) $ \\
		\hline \hline
		ET & 0 &$C_{e \tau}$ &\ 0 & \ 1 &\ 0.1 &\ 1&$Y_T = f\left(\begin{matrix}1&0&C_{e \tau}\\0&0.1&0\\ 0&0&\sqrt{1-C_{e \tau}^2}\\\end{matrix}\right) $ \\
		\hline \hline
		EM & 0 & 0 & $C_{e\mu}$&1&1&0.1& $Y_T = f\left(\begin{matrix}\sqrt{1-C_{e\mu}^2}&0&0\\ C_{e\mu}&1&0\\0&0&0.1\\\end{matrix}\right) $ \\
		\hline
	\end{tabular}
\end{table}

\subsection{$\BR(Z\to l_i l_j)$}

In this section, we study the variation of $\BR(Z\to l_i l_j)$ after applying constraints from $\BR(l_i\to l_j\gamma)$. The decays $l_i\to l_j\gamma$ have not been observed in any experiment. Therefore, the branching ratios for these decays are constrained as shown in table~\ref{tab:rad-lfv}.

\begin{table}[htbp]
	\centering
	\caption{Experimental upper limits for lepton-flavor-violating radiative decays, $l_i\to l_j\gamma$.}
	\label{tab:rad-lfv}
	\begin{tabular}{|l|c|c|c|c|}
		\hline
		Decay & Upper Limit & Experiment\\
		\hline \hline
		$BR(\tau \longrightarrow \mu \gamma )$& $4.2\times{10}^{-8}$ \cite{ref62} &Belle\\
		\hline
		$BR(\tau \longrightarrow e \gamma )$& $5.6\times{10}^{-8}$ \cite{ref62} &Belle\\
		\hline
		$BR(\mu \longrightarrow e \gamma )$& $4.2\times{10}^{-13}$ \cite{ref63} &MEG\\
		\hline
	\end{tabular}
\end{table}

\begin{figure}[htbp]
	\centering
	\includegraphics[width=0.8\textwidth]{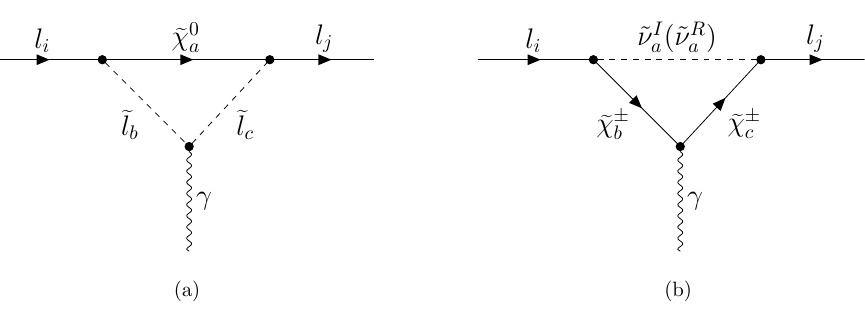}
	\caption{One-loop Feynman diagrams contributing to $\BR(l_i\to l_j\gamma)$ in the MSSM-seesaw type-II model.}
	\label{fig:ligamma-diagrams}
\end{figure}

In supersymmetry, LFV processes can originate from the non-diagonal elements of the left-slepton mass matrix. Therefore, in the mass-insertion method with the leading-logarithm approximation, the branching ratios of $l_i\to l_j\gamma$ decays can be approximated as \cite{ref37,ref41}
\begin{equation}
	\BR(l_i\to l_j\gamma)\propto
	\alpha^3 m_{l_i}^5
	\frac{\left|\Delta m_{\tilde{L}_{ij}}^2\right|^2}{\tilde{m}^8}
	\tan^2\beta .
	\label{eq:mi-ligamma}
\end{equation}
Here $\alpha$ is the electroweak coupling constant, $m_{l_i}$ is the mass of lepton $i$, and $\tilde{m}$ is the average SUSY mass entering the loops. From eqs.~\eqref{eq:rge-lfv} and \eqref{eq:mi-ligamma}, $\BR(l_i\to l_j\gamma)$ is mainly related to the SUSY mass scale and to the off-diagonal elements of the left-slepton mass matrix. These off-diagonal elements are almost completely governed by the choice of soft SUSY-breaking parameters in the heavy seesaw sector. Furthermore, $\BR(l_i\to l_j\gamma)$ is proportional to $\tan^2\beta$, which leads to larger LFV branching ratios. Therefore, we fix $A_0=0~\GeV$ (to exclude trilinear couplings) and $\tan\beta=5$ in all calculations. We also fix $m_0=550~\GeV$ and $m_{1/2}=200~\GeV$ by considering the SUSY-particle mass limits in table~\ref{tab:sparticle-mass}. We use the LFV decays $l_i\to l_j\gamma$ to constrain the parameter $f$, while the values of the $Y_T$ matrix are fixed as shown in table~\ref{tab:yukawa-scenarios}. Sparticle-mediated diagrams (sleptons, sneutrinos, neutralinos, and charginos) for the LFV decays $l_i\to l_j\gamma$ in the MSSM-seesaw type-II model are shown in figure~\ref{fig:ligamma-diagrams}.

The off-shell amplitude for $l_i\to l_j\gamma$ is \cite{ref45,ref64}
\begin{equation}
	\mathcal{M}=e\epsilon_\alpha^*\bar{u}_i(p-q)
	\left[
	q^2\gamma^\alpha(K_1^L P_L+K_1^R P_R)
	+m_{l_i}i\sigma^{\alpha\beta}q_\beta(K_2^L P_L+K_2^R P_R)
	\right]u_j(p).
	\label{eq:ligamma-amplitude}
\end{equation}
Here $q$ is the photon momentum, $e$ is the electric charge, and $\epsilon_\alpha$ is the photon polarization vector. The spinors $u_i$ and $u_j$ denote the wave functions of the antilepton/lepton, and $p$ is the momentum of lepton $l_i$. In the limit $q\to0$, the analytic expression for the branching ratio of the $l_i\to l_j\gamma$ decays is
\begin{equation}
	\BR(l_i\to l_j\gamma)=
	\alpha\frac{m_{l_i}^5}{\Gamma_{l_i}}\left(|K_2^L|^2+|K_2^R|^2\right).
	\label{eq:ligamma-br}
\end{equation}
Here $\alpha$ is the fine-structure constant and $\Gamma_{l_i}$ is the total decay width of $l_i$. The coefficients $K_2^{L/R}$ are combinations of the coefficients corresponding to the Feynman diagrams shown in figure~\ref{fig:ligamma-diagrams} and can be written as
\begin{equation}
	K_2^{L/R}=K_{2a}^{L/R}+K_{2b}^{L/R}.
	\label{eq:k2-sum}
\end{equation}
The contributions from neutralino-slepton loops and chargino-sneutrino loops are shown in figures~\ref{fig:ligamma-diagrams}(a) and \ref{fig:ligamma-diagrams}(b), respectively. The contributions of figure~\ref{fig:ligamma-diagrams}(a) are
\begin{align}
	K_2^L &=
	2V_\gamma\left[
	V_1^L V_2^R(C_2+C_{12}+C_{11})m_{l_i}
	+V_1^R V_2^L(C_1+C_{12}+C_{11})m_{l_j}
	- V_1^R V_2^R(C_0+C_1+C_2)m_{\tilde{\chi}_a^0}
	\right],
	\label{eq:k2-neutralino}\\
	K_2^R&=K_2^L(L\leftrightarrow R).
	\label{eq:k2-neutralino-r}
\end{align}
The couplings corresponding to figure~\ref{fig:ligamma-diagrams}(a) are
\[
V_\gamma=\Gamma_{\tilde{l}_c\tilde{l}_b^{*}\gamma} \quad,\quad
V_1^{L/R}=\Gamma_{\bar{l}_i\tilde{\chi}_a^{0}\tilde{l}_b}^{L/R} \quad,\quad
V_2^{L/R}=\Gamma_{\tilde{\chi}_a^{0} l_j \tilde{l}_c^{*}}^{L/R}.
\]
Here $\Gamma_{\tilde{l}_c\tilde{l}_b^{*}\gamma}$ is the slepton-slepton-photon coupling, $\Gamma_{\bar{l}_i\tilde{\chi}_a^{0}\tilde{l}_b}^{L/R}$ is the left(right)-handed antilepton-neutralino-slepton coupling, and $\Gamma_{\tilde{\chi}_a^{0} l_j \tilde{l}_b^{*}}^{L/R}$ is the left(right)-handed neutralino-lepton-slepton coupling. Their explicit forms are shown in appendix~\ref{sec:appendix}.

For figure~\ref{fig:ligamma-diagrams}(b), the contributions are
\begin{align}
	K_2^L &=
	-2(V_\gamma^L)^*V_1^L V_2^R C_{12}m_{l_i}
	+2(V_\gamma^R)^*V_1^R V_2^L(C_2+C_{12}+C_{22})m_{l_j}
	\nonumber\\
	&\quad
	+2(V_\gamma^L)^*V_1^R V_2^R C_1m_{\tilde{\chi}_b^-}
	-2(V_\gamma^R)^*V_1^R V_2^R(C_0+C_1+C_2)m_{\tilde{\chi}_c^-},
	\label{eq:k2-chargino}\\
	K_2^R&=K_2^L(L\leftrightarrow R).
	\label{eq:k2-chargino-r}
\end{align}
The couplings corresponding to figure~\ref{fig:ligamma-diagrams}(b) are
\[
V_\gamma^{L/R}=\Gamma_{\tilde{\chi}_b^{+}\tilde{\chi}_c^{-}\gamma}^{L/R} \quad,\quad
V_1^{L/R}=\Gamma_{\bar{l}_i\tilde{\chi}_b^{-}\tilde{\nu}_a^{I}}^{L/R} \quad,\quad
V_2^{L/R}=\Gamma_{\tilde{\chi}_c^{+} l_j \tilde{\nu}_a^{I}}^{L/R}.
\]
Here $\Gamma_{\tilde{\chi}_b^{+}\tilde{\chi}_c^{-}\gamma}^{L/R}$ represents the left(right)-handed chargino-chargino-photon coupling, $\Gamma_{\bar{l}_i\tilde{\chi}_b^{-}\tilde{\nu}_a^{I}}^{L/R}$ is the left(right)-handed antilepton-chargino-CP-odd sneutrino coupling, and $\Gamma_{\tilde{\chi}_c^{+} l_j \tilde{\nu}_a^{I}}^{L/R}$ is the left(right)-handed chargino-lepton-CP-odd sneutrino coupling. Their explicit forms are given in appendix~\ref{sec:appendix}. For CP-even sneutrino couplings, we replace $\tilde{\nu}^I$ by $\tilde{\nu}^R$ in the last two vertices, $V_1^{L/R}$ and $V_2^{L/R}$, and in the arguments of the $C$ functions.

\begin{figure}[htbp]
	\centering
	\includegraphics[width=0.95\textwidth]{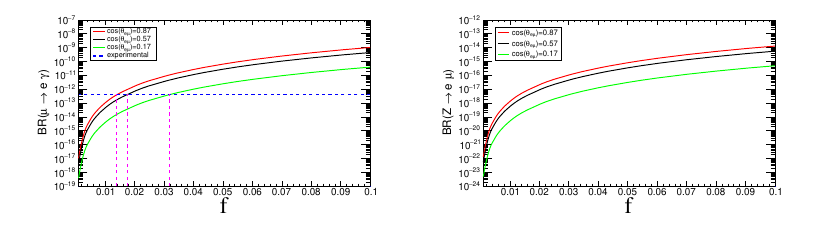}
	\caption{$\BR(\mu\to e\gamma)$ versus $f$ (left) and $\BR(Z\to e\mu)$ versus $f$ (right). In both plots, $\cos(\theta_{e\mu})=0.87$ is shown by the red line, $\cos(\theta_{e\mu})=0.57$ by the black line, and $\cos(\theta_{e\mu})=0.17$ by the green line.}
	\label{fig:muegamma}
\end{figure}

After studying the analytic expression of $\BR(l_i\to l_j\gamma)$, we begin the numerical discussion with $\mu$ decays because their experimental limits are the strongest. For this analysis, the parameters are fixed as $m_0=550~\GeV$, $m_{1/2}=200~\GeV$, $\tan\beta=5$, $M_T=5\times10^{13}~\GeV$, and $A_0=0~\GeV$. We plot both $\BR(\mu\to e\gamma)$ and $\BR(Z\to e\mu)$ versus $f$ in figure~\ref{fig:muegamma} for $\cos(\theta_{e\mu})=0.87$, 0.57, and 0.17. The horizontal dotted line is the current experimental limit on $\BR(\mu\to e\gamma)$ shown in table~\ref{tab:rad-lfv}.

Both $\BR(\mu\to e\gamma)$ and $\BR(Z\to e\mu)$ increase as $f$ varies from 0.001 to 0.1. The prediction for $\BR(\mu\to e\gamma)$ exceeds the experimental limit at $f=0.014$, 0.0175, and 0.031 for the red, black, and green lines, respectively. In this case, $\BR(Z\to e\mu)$ is $4.75\times10^{-18}$, $5.12\times10^{-18}$, and $5.09\times10^{-18}$ at $f=0.014$ for $\cos(\theta_{e\mu})=0.87$, at $f=0.0175$ for $\cos(\theta_{e\mu})=0.57$, and at $f=0.031$ for $\cos(\theta_{e\mu})=0.17$, respectively. Thus, under the constraints from $\BR(\mu\to e\gamma)$, $\BR(Z\to e\mu)$ is about $5\times10^{-18}$. This predicted value is eleven orders of magnitude below the current experimental limit and eight orders below the sensitivity of future colliders shown in table~\ref{tab:zlfv-limits}.

\begin{figure}[htbp]
	\centering
	\includegraphics[width=0.95\textwidth]{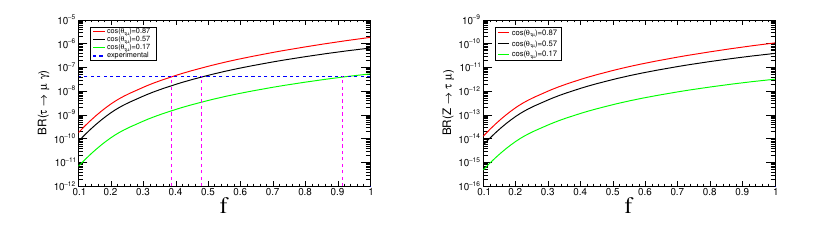}
	\caption{$\BR(\tau\to\mu\gamma)$ versus $f$ (left) and $\BR(Z\to\tau\mu)$ versus $f$ (right). In both plots, $\cos(\theta_{\tau\mu})=0.87$ is shown by the red line, $\cos(\theta_{\tau\mu})=0.57$ by the black line, and $\cos(\theta_{\tau\mu})=0.17$ by the green line.}
	\label{fig:taumugamma}
\end{figure}

In figure~\ref{fig:taumugamma}, we plot both $\BR(\tau\to\mu\gamma)$ and $\BR(Z\to\tau\mu)$ versus $f$ for $\cos(\theta_{\tau\mu})=0.87$, 0.57, and 0.17. The horizontal dotted line represents the current experimental limit on $\BR(\tau\to\mu\gamma)$ shown in table~\ref{tab:rad-lfv}.

Both $\BR(\tau\to\mu\gamma)$ and $\BR(Z\to\tau\mu)$ increase as $f$ varies from 0.1 to 1. The prediction for $\BR(\tau\to\mu\gamma)$ exceeds the current experimental limit at $f=0.387$, 0.48, and 0.915 for the red, black, and green lines, respectively. In this case, $\BR(Z\to\tau\mu)$ is $2.81\times10^{-12}$, $2.63\times10^{-12}$, and $2.37\times10^{-12}$ at $f=0.387$ for $\cos(\theta_{\tau\mu})=0.87$, at $f=0.48$ for $\cos(\theta_{\tau\mu})=0.57$, and at $f=0.915$ for $\cos(\theta_{\tau\mu})=0.17$, respectively. Hence, under the constraints from $\tau\to\mu\gamma$, $\BR(Z\to\tau\mu)$ is approximately $2\times10^{-12}$. This predicted value is six orders of magnitude below the current experimental limit and three orders below the sensitivity of future colliders shown in table~\ref{tab:zlfv-limits}.

\begin{figure}[htbp]
	\centering
	\includegraphics[width=0.95\textwidth]{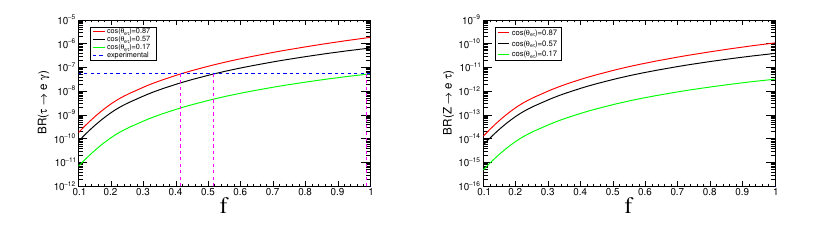}
	\caption{$\BR(\tau\to e\gamma)$ versus $f$ (left) and $\BR(Z\to e\tau)$ versus $f$ (right). In both plots, $\cos(\theta_{e\tau})=0.87$ is shown by the red line, $\cos(\theta_{e\tau})=0.57$ by the black line, and $\cos(\theta_{e\tau})=0.17$ by the green line.}
	\label{fig:tauegamma}
\end{figure}

Similarly, figure~\ref{fig:tauegamma} shows both $\BR(\tau\to e\gamma)$ and $\BR(Z\to e\tau)$ versus $f$ for $\cos(\theta_{e\tau})=0.87$, 0.57, and 0.17. The horizontal dotted line is the current experimental limit on $\BR(\tau\to e\gamma)$ shown in table~\ref{tab:rad-lfv}.

Both $\BR(\tau\to e\gamma)$ and $\BR(Z\to e\tau)$ increase as $f$ varies from 0.1 to 1. The prediction for $\BR(\tau\to e\gamma)$ exceeds the current experimental limit at $f=0.415$, 0.515, and 0.99 for the red, black, and green lines, respectively. In this case, $\BR(Z\to e\tau)$ is $3.46\times10^{-12}$, $3.44\times10^{-12}$, and $3.05\times10^{-12}$ at $f=0.415$ for $\cos(\theta_{e\tau})=0.87$, at $f=0.515$ for $\cos(\theta_{e\tau})=0.57$, and at $f=0.99$ for $\cos(\theta_{e\tau})=0.17$, respectively.
Thus, under the constraints from $\tau\to e\gamma$, $\BR(Z\to e\tau)$ is about $3\times10^{-12}$. This predicted value is six orders of magnitude below the current experimental limit and three orders below the sensitivity of future colliders shown in table~\ref{tab:zlfv-limits}.

The numerical results indicate that, as $\cos(\theta_{ij})$ increases, the allowed values of $f$ become smaller, while the predicted values of $\BR(Z\to l_i l_j)$ are approximately equal for $\cos(\theta_{ij})=0.87$, 0.57, and 0.17.

\begin{figure}[htbp]
	\centering
	\includegraphics[width=0.95\textwidth]{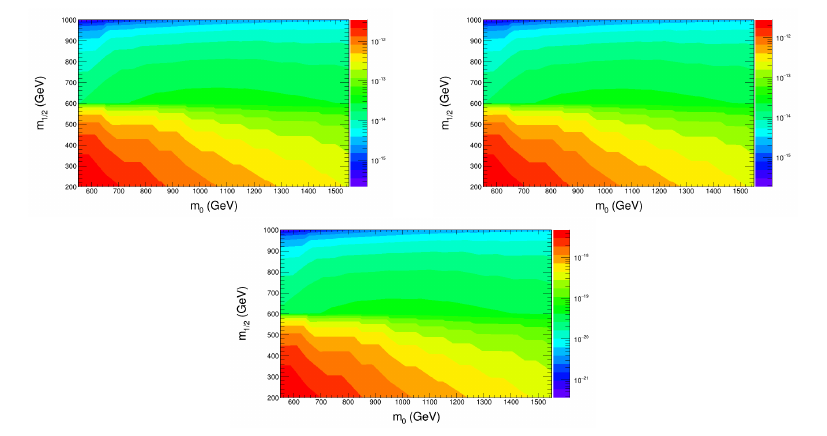}
	\caption{$\BR(Z\to l_i l_j)$ as a contour in the $m_0$ and $m_{1/2}$ plane at $\cos(\theta_{ij})=0.87$. Here $f=0.387$ for $\BR(Z\to\mu\tau)$(top-right), $f=0.415$ for $\BR(Z\to e\tau)$(top-left), and $f=0.014$ for $\BR(Z\to e\mu)$(bottom).}
	\label{fig:costheta30}
\end{figure}

\begin{figure}[htbp]
	\centering
	\includegraphics[width=0.95\textwidth]{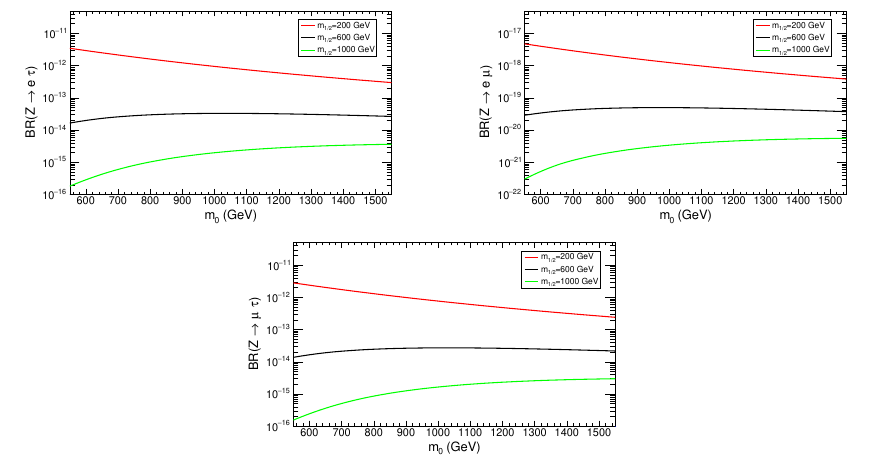}
	\caption{$\BR(Z\to l_i l_j)$ as a function of $m_0$ at $m_{1/2}=200$, 600, and $1000~\GeV$. Here $f=0.387$ for $\BR(Z\to\mu\tau)$, $f=0.415$ for $\BR(Z\to e\tau)$, and $f=0.014$ for $\BR(Z\to e\mu)$ at $\cos(\theta_{ij})=0.87$.}
	\label{fig:m0-constraints}
\end{figure}

\begin{figure}[htbp]
	\centering
	\includegraphics[width=0.95\textwidth]{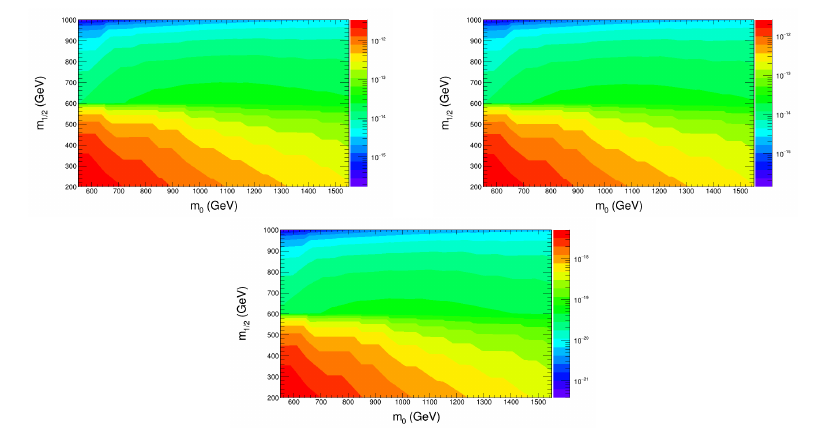}
	\caption{$\BR(Z\to l_i l_j)$ as a contour in the $m_0$ and $m_{1/2}$ plane at $\cos(\theta_{ij})=0.57$. Here $f=0.48$ for $\BR(Z\to\mu\tau)$(top-right), $f=0.515$ for $\BR(Z\to e\tau)$(top-left), and $f=0.0175$ for $\BR(Z\to e\mu)$(bottom).}
	\label{fig:costheta55}
\end{figure}

\begin{figure}[htbp]
	\centering
	\includegraphics[width=0.95\textwidth]{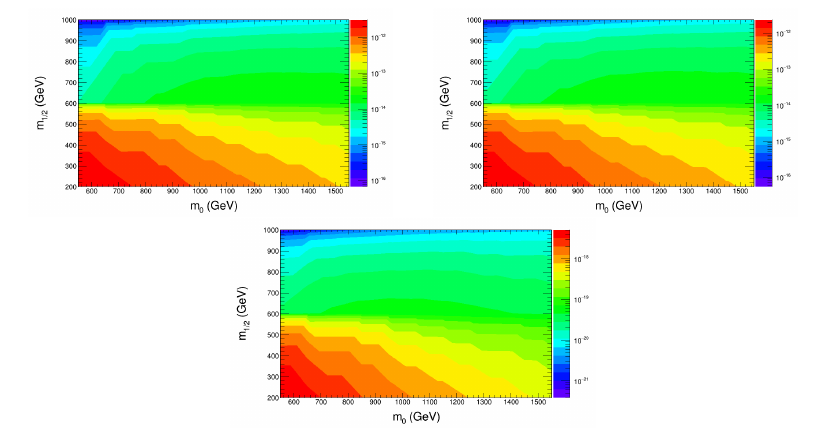}
	\caption{$\BR(Z\to l_i l_j)$ as a contour in the $m_0$ and $m_{1/2}$ plane at $\cos(\theta_{ij})=0.17$. Here $f=0.915$ for $\BR(Z\to\mu\tau)$(top-right), $f=0.99$ for $\BR(Z\to e\tau)$(top-left), and $f=0.031$ for $\BR(Z\to e\mu)$(bottom).}
	\label{fig:costheta80}
\end{figure}

The parameters $m_0$ and $m_{1/2}$ are associated with the soft-breaking slepton mass matrices and soft-breaking gaugino masses at the GUT scale. As mentioned earlier, $m_0$ appears in the off-diagonal elements of the slepton mass matrix and therefore affects LFV. Figures~\ref{fig:costheta30}, \ref{fig:costheta55}, and \ref{fig:costheta80} show the variation of $\BR(Z\to l_i l_j)$ as a contour in the $m_0$ and $m_{1/2}$ space at $\cos(\theta_{ij})=0.87$, 0.57, and 0.17 respectively. Whereas Figure~\ref{fig:m0-constraints} shows the variation of $\BR(Z\to l_i l_j)$ as a function of $m_0$ for $m_{1/2}=200$, 600, and $1000~\GeV$


The numerical results show that when both $m_0$ and $m_{1/2}$ increase, the $Z$-decay branching ratios decrease. This indicates that large values of $m_0$ and $m_{1/2}$ suppress LFV in $Z$ decays. The best values of BRs appear in the red region, corresponding to $m_0$ in the range [550,720] GeV and $m_{1/2}$ in the range [200, 340] GeV as shown in Figures~\ref{fig:costheta30}, \ref{fig:costheta55}, and \ref{fig:costheta80}.Thus, ZLFV decays can occur at the low mass limit of SUSY particles, whose masses depend directly on these parameters.

In addition, the upper two plots which corresponding to $\tau$ channels ($e\tau$ and $\mu\tau$) are almost identical in their contour geometry and magnitude. This behavior reflects the dominance of the large third-generation Yukawa couplings in ZLFV decays.
Furthermore, the bottom plot ($e\mu$ channel) shows a significantly suppressed branching ratio. This reflects the incredibly stringent experimental constraints on $\mu \to e\gamma$ decays, which force the theoretical parameter space for $e\mu$ mixing to be highly constrained compared to $\tau$ transitions.

%
Besides, under the constraints from $l_i\to l_j\gamma$ decays and with $\cos(\theta_{ij})=0.87$, the numerical results indicate that the branching ratios for $Z\to e\mu$, $Z\to e\tau$, and $Z\to\mu\tau$ are within the ranges $[3.00\times10^{-22},4.75\times10^{-18}]$, $[1.82\times10^{-16},3.46\times10^{-12}]$, and $[1.55\times10^{-16},2.81\times10^{-12}]$, respectively.

While with $\cos(\theta_{ij})=0.57$, the branching ratios for $Z\to e\mu$, $Z\to e\tau$, and $Z\to\mu\tau$ are within the ranges $[3.24\times10^{-22},5.12\times10^{-18}]$, $[1.39\times10^{-16},3.44\times10^{-12}]$, and $[1.17\times10^{-16},2.63\times10^{-12}]$, respectively.

While with $\cos(\theta_{ij})=0.17$, the branching ratios for $Z\to e\mu$, $Z\to e\tau$, and $Z\to\mu\tau$ are within the ranges $[3.20\times10^{-22},5.09\times10^{-18}]$, $[5.70\times10^{-17},3.05\times10^{-12}]$, and $[4.45\times10^{-17},2.37\times10^{-12}]$, respectively.

The best predictions of the branching ratios of LFV $Z$ decays are in table~\ref{tab:BRvalues}
\begin{table}[htbp]
	\centering
	\caption{Upper-limit values for the branching ratios of LFV $Z$ decays.}
	\label{tab:BRvalues}
	\begin{tabular}{|l|c|c|c|c|}
		\hline
		Z Decays & 	BR$(Z\longrightarrow \mu\tau)$ &BR$(Z\longrightarrow e\tau)$& BR$(Z\longrightarrow e \mu)$\\
		\hline
		\hline
		cos($\theta_{ij})$=0.87 &$2.81\times{10}^{-12}$& $3.46\times10^{-12}$& $4.75\times10^{-18}$\\
		\hline
		cos($\theta_{ij})$=0.57 &$2.63\times{10}^{-12}$& $3.44\times10^{-12}$& $5.12\times10^{-18}$\\
		\hline
		cos($\theta_{ij})$=0.17 &$2.37\times{10}^{-12}$& $3.05\times10^{-12}$& $5.09\times10^{-18}$\\
		\hline
	\end{tabular}
\end{table} 

		\section{Conclusions}

This article studies LFV $Z$-boson decays, $Z\to l_i l_j$, in the CMSSM-seesaw type-II model. In supersymmetric models, the type-II seesaw mechanism can be realized by adding a scalar triplet superfield. After deriving the analytic expressions for the branching ratios of both $Z$-boson and radiative two-body $l_i\to l_j\gamma$ decays, we predicted the branching-ratio values for LFV $Z$-boson decays in the studied model. The numerical study was performed by imposing the following constraints: $R$ parity is conserved, small neutrino masses are fitted, and the neutralino is the lightest supersymmetric particle of the considered model. In addition, the sparticle masses (charginos, sleptons, sneutrinos, and neutralinos) are required to be above the recent experimental mass limits. We also imposed constraints from $\BR(l_i\to l_j\gamma)$ and perturbativity limits on the model parameters. To investigate the small masses of left-handed neutrinos, the scalar-triplet mass $M_T$ must be of order $\geq5\times10^{13}~\GeV$.

After satisfying all constraints on the model parameters, but without constraints from the $l_i\to l_j\gamma$ decays, the maximum values of the LFV $Z$-boson decay branching ratios are summarized in table~\ref{tab:results}. We also find that the parameters $f$, $\cos(\theta_{ij})$, and $A_0$ enhance LFV in $Z$-boson decays, while $m_0$ and $m_{1/2}$ suppress $Z$ LFV.

After considering the constraints from $l_i\to l_j\gamma$ decays on the parameter space of $f$, $\BR(Z\to\mu\tau)$ and $\BR(Z\to e\tau)$ can be of order $10^{-12}$, while $\BR(Z\to e\mu)$ can be of order $10^{-18}$.

\begin{table}[htbp]
	\centering
	\caption{Upper-limit values for the branching ratios of LFV $Z$ decays. The table compares the predictions of this study with ATLAS limits and FCC-ee/CEPC expectations. ``Our results denotes results with constraints from $l_i\to l_j\gamma$ decays.}
	\label{tab:results}
	\begin{tabular}{|l|c|c|c|c|}
		\hline
		Z Decays & Our results &LHC(90\%)& FCC-ee/CEPC\\
		\hline
		\hline
		BR$(Z\longrightarrow \mu\tau)$ &$2.00\times{10}^{-12}$& $7.20\times{10}^{-6}$& ${10}^{-9}$\\
		\hline
		BR$(Z\longrightarrow e\tau)$  &$3.00\times{10}^{-12}$& $7.00\times{10}^{-6}$ & ${10}^{-9}$\\
		\hline
		BR$(Z\longrightarrow e \mu)$ &$5.00\times{10}^{-18}$& $2.62\times{10}^{-7}$ & $10^{-8}-{10}^{-10}$\\
		\hline
	\end{tabular}
\end{table}

These values are outside the experimental upper limits for the LHC, while they coincide with the sensitivities of future colliders (FCC-ee/CEPC) for the scenario without constraints from $l_i\to l_j\gamma$ decays, as shown in table~\ref{tab:results}. On the other hand, in the scenario with constraints from $l_i\to l_j\gamma$ decays on the parameter space of $f$, the above-mentioned results for LFV $Z$-boson decay branching ratios receive an additional suppression of about $10^{-3}$ for $\BR(Z\to l\tau)$ and $10^{-8}$ for $\BR(Z\to e\mu)$ compared with the sensitivity of future colliders (FCC-ee/CEPC). Hence, the predicted branching ratios are several orders of magnitude below the recent experimental limits for both scenarios, implying a very low possibility of observing LFV $Z$-boson decays in future collider experiments. Finally, based on this study, many analyses of LFV $Z$-boson decays can be performed in other BSM and supersymmetric models by using this method to predict branching ratios for LFV observables in different decay modes.

%
%
%


	\acknowledgments
	
	The authors would like to thank the administration of scientific research at Erzincan Binali Yildirim University/Türkiye, Al-Furat and Idlib University/Syria for supporting and funding this work.
	
	\appendix
	\section{Appendix: Vertices}
\label{sec:appendix}

\subsection{Two fermions--one scalar interactions}

\begin{align}
	\Gamma_{\tilde{\chi}_i^0 l_j\tilde{l}_k^*}^{L}
	&=i\left[
	\frac{1}{\sqrt{2}}g_1N_{i1}^*\sum_{a=1}^{3}U_{L,ja}^{l,*}Z_{ka}^{E}
	+\frac{1}{\sqrt{2}}g_2N_{i2}^*\sum_{a=1}^{3}U_{L,ja}^{l,*}Z_{ka}^{E}
	-N_{i3}^*\sum_{b=1}^{3}\sum_{a=1}^{3}U_{L,jb}^{l,*}Y_{l,ab}Z_{k,3+a}^{E}
	\right],
	\label{eq:a9}\\
	\Gamma_{\tilde{\chi}_i^0 l_j\tilde{l}_k^*}^{R}
	&=i\left[
	-\sqrt{2}g_1\sum_{a=1}^{3}Z_{k,3+a}^{E}U_{R,ja}^{l}N_{i1}
	-\sum_{b=1}^{3}\sum_{a=1}^{3}Y_{l,ab}^*U_{R,ja}^{l}Z_{kb}^{E}N_{i3}
	\right],
	\label{eq:a10}\\
	\Gamma_{\bar{l}_j\tilde{\chi}_i^0\tilde{l}_k}^{L}
	&=i\left[
	-N_{j3}^*\sum_{b=1}^{3}\sum_{a=1}^{3}Z_{kb}^{E,*}U_{R,ia}^{l,*}Y_{l,ab}
	-\sqrt{2}g_1N_{j1}^*\sum_{a=1}^{3}Z_{k,3+a}^{E,*}U_{R,ia}^{l,*}
	\right],
	\label{eq:a11}\\
	\Gamma_{\bar{l}_j\tilde{\chi}_i^0\tilde{l}_k}^{R}
	&=i\left[
	\frac{1}{\sqrt{2}}\sum_{a=1}^{3}Z_{ka}^{E,*}U_{L,ia}^{l}
	\left(g_1N_{j1}+g_2N_{j2}\right)
	-\sum_{b=1}^{3}\sum_{a=1}^{3}Y_{l,ab}^*Z_{k,3+a}^{E,*}U_{L,ib}^{l}N_{j3}
	\right],
	\label{eq:a12}\\
	\Gamma_{\bar{l}_i\tilde{\chi}_j^-\tilde{\nu}_k^I}^{L}
	&=-\frac{1}{\sqrt{2}}U_{j2}^*
	\sum_{b=1}^{3}Z_{kb}^{I,*}\sum_{a=1}^{3}U_{R,ia}^{l,*}Y_{l,ab}^{l,*},
	\label{eq:a13}\\
	\Gamma_{\bar{l}_i\tilde{\chi}_j^-\tilde{\nu}_k^I}^{R}
	&=\frac{1}{4}\left[
	2\sqrt{2}g_2\sum_{a=1}^{3}Z_{ka}^{I,*}U_{L,ia}^{l}V_{j1}
	+v_u\left(
	\sum_{b=1}^{3}Z_{kb}^{I,*}\sum_{a=1}^{3}\kappa_{\nu,ab}^*U_{L,ia}^{l}
	+\sum_{b=1}^{3}\sum_{a=1}^{3}\kappa_{\nu,ab}^*Z_{ka}^{I,*}U_{L,ib}^{l}
	\right)V_{j2}
	\right],
	\label{eq:a14}\\
	\Gamma_{\bar{l}_i\tilde{\chi}_j^-\tilde{\nu}_k^R}^{L}
	&=\frac{i}{\sqrt{2}}U_{j2}^*
	\sum_{b=1}^{3}Z_{kb}^{R,*}\sum_{a=1}^{3}U_{R,ia}^{l,*}Y_{l,ab}^{l,*},
	\label{eq:a15}\\
	\Gamma_{\bar{l}_i\tilde{\chi}_j^-\tilde{\nu}_k^R}^{R}
	&=\frac{i}{4}\left[
	-2\sqrt{2}g_2\sum_{a=1}^{3}Z_{ka}^{R,*}U_{L,ia}^{l}V_{j1}
	+v_u\left(
	\sum_{b=1}^{3}Z_{kb}^{R,*}\sum_{a=1}^{3}\kappa_{\nu,ab}^*U_{L,ia}^{l}
	+\sum_{b=1}^{3}\sum_{a=1}^{3}\kappa_{\nu,ab}^*Z_{ka}^{R,*}U_{L,ib}^{l}
	\right)V_{j2}
	\right],
	\label{eq:a16}\\
	\Gamma_{\tilde{\chi}_i^+l_j\tilde{\nu}_k^I}^{L}
	&=\frac{1}{4}\left[
	-\sqrt{2}\,2g_2V_{i1}^*\sum_{a=1}^{3}U_{L,ja}^{l,*}Z_{ka}^{I,*}
	-v_uV_{i2}^*\left(
	\sum_{b=1}^{3}Z_{kb}^{I,*}\sum_{a=1}^{3}U_{L,ja}^{l,*}\kappa_{\nu,ab}
	+\sum_{b=1}^{3}\sum_{a=1}^{3}U_{L,jb}^{l,*}Z_{ka}^{I,*}\kappa_{\nu,ab}
	\right)\right],
	\label{eq:a17}\\
	\Gamma_{\tilde{\chi}_i^+l_j\tilde{\nu}_k^I}^{R}
	&=\frac{1}{\sqrt{2}}\sum_{b=1}^{3}Z_{kb}^{I,*}
	\sum_{a=1}^{3}Y_{l,ab}^{l,*}U_{R,ja}^{l}U_{i2},
	\label{eq:a18}\\
	\Gamma_{\tilde{\chi}_i^+l_j\tilde{\nu}_k^R}^{L}
	&=\frac{i}{4}\left[
	-2\sqrt{2}g_2V_{i1}^*\sum_{a=1}^{3}U_{L,ja}^{l,*}Z_{ka}^{R,*}
	+v_uV_{i2}^*\left(
	\sum_{b=1}^{3}Z_{kb}^{R,*}\sum_{a=1}^{3}U_{L,ja}^{l,*}\kappa_{\nu,ab}
	+\sum_{b=1}^{3}\sum_{a=1}^{3}U_{L,jb}^{l,*}Z_{ka}^{R,*}\kappa_{\nu,ab}
	\right)\right],
	\label{eq:a19}\\
	\Gamma_{\tilde{\chi}_i^+l_j\tilde{\nu}_k^R}^{R}
	&=\frac{i}{\sqrt{2}}\sum_{b=1}^{3}Z_{kb}^{R,*}
	\sum_{a=1}^{3}Y_{l,ab}^{l,*}U_{R,ja}^{l}U_{i2}.
	\label{eq:a20}
\end{align}

\subsection{Two fermions--$Z$-boson interactions}
\begin{align}
	\Gamma_{\tilde{\chi}_i^0\tilde{\chi}_j^0 Z}^{L}
	&=-\frac{i}{2}\left(g_1\sin\Theta_W+g_2\cos\Theta_W\right)
	\left(N_{j3}^*N_{i3}-N_{j4}^*N_{i4}\right),
	\label{eq:a1}\\
	\Gamma_{\tilde{\chi}_i^0\tilde{\chi}_j^0 Z}^{R}
	&=+\frac{i}{2}\left(g_1\sin\Theta_W+g_2\cos\Theta_W\right)
	\left(N_{i3}N_{j3}^*-N_{i4}N_{j4}^*\right),
	\label{eq:a2}\\
	\Gamma_{\tilde{\chi}_i^+\tilde{\chi}_j^- Z}^{L}
	&=\frac{i}{2}\left[
	2g_2U_{j1}^*\cos\Theta_W U_{i1}
	+U_{j2}^*(-g_1\sin\Theta_W+g_2\cos\Theta_W)U_{i2}
	\right],
	\label{eq:a3}\\
	\Gamma_{\tilde{\chi}_i^+\tilde{\chi}_j^- Z}^{R}
	&=\frac{i}{2}\left[
	2g_2V_{i1}^*\cos\Theta_W V_{j1}
	+V_{i2}^*(-g_1\sin\Theta_W+g_2\cos\Theta_W)V_{j2}
	\right],
	\label{eq:a4}\\
	\Gamma_{\bar{l}_i l_j Z}^{L}
	&=\frac{i}{2}\delta_{ij}\left(-g_1\sin\Theta_W+g_2\cos\Theta_W\right),
	\label{eq:a5}\\
	\Gamma_{\bar{l}_i l_j Z}^{R}
	&=-ig_1\delta_{ij}\sin\Theta_W .
	\label{eq:a6}
\end{align}

\subsection{Two scalars--$Z$-boson interactions}
\begin{align}
	\Gamma_{\tilde{\nu}_i^I\tilde{\nu}_j^R Z}
	&=\frac{1}{2}\left(g_1\sin\Theta_W+g_2\cos\Theta_W\right)
	\sum_{a=1}^{3} Z_{ia}^{I,*}Z_{ja}^{R,*},
	\label{eq:a7}\\
	\Gamma_{\tilde{l}_i\tilde{l}_j^* Z}
	&=\frac{i}{2}\left[
	-2g_1\sin\Theta_W\sum_{a=1}^{3}Z_{i,3+a}^{E,*}Z_{j,3+a}^{E}
	+\left(-g_1\sin\Theta_W+g_2\cos\Theta_W\right)
	\sum_{a=1}^{3}Z_{ia}^{E,*}Z_{ja}^{E}
	\right].
	\label{eq:a8}
\end{align}

\subsection{Two scalars--$\gamma$ interactions}

\begin{equation}
	\Gamma_{\tilde{l}_i\tilde{l}_j^*\gamma}
	=\frac{i}{2}\left[
	2g_1\cos\Theta_W\sum_{a=1}^{3}Z_{i,3+a}^{E,*}Z_{j,3+a}^{E}
	+\left(g_1\cos\Theta_W+g_2\sin\Theta_W\right)
	\sum_{a=1}^{3}Z_{ia}^{E,*}Z_{ja}^{E}
	\right].
	\label{eq:a21}
\end{equation}

\subsection{Two fermions--$\gamma$ interactions}

\begin{align}
	\Gamma_{\tilde{\chi}_i^+\tilde{\chi}_j^-\gamma}^{L}
	&=\frac{i}{2}\left[
	2g_2U_{j1}^*\sin\Theta_W U_{i1}
	+U_{j2}^*\left(g_1\cos\Theta_W+g_2\sin\Theta_W\right)U_{i2}
	\right],
	\label{eq:a22}\\
	\Gamma_{\tilde{\chi}_i^+\tilde{\chi}_j^-\gamma}^{R}
	&=\frac{i}{2}\left[
	2g_2V_{i1}^*\sin\Theta_W V_{j1}
	+V_{i2}^*\left(g_1\cos\Theta_W+g_2\sin\Theta_W\right)V_{j2}
	\right].
	\label{eq:a23}
\end{align}

	
	\bibliographystyle{JHEP}
	\bibliography{biblio.bib}

@article{ref1,
author = {Primulandoando, R. and Julio, J.  and Uttayarat, P.},
title = {Scalar phenomenology in type-II seesaw model},
journal = {J. High Energ. Phys.},
volume = {02},
year = {2022},
issn = {1029-8479},
doi = {10.1007/JHEP08(2019)024},
eprint={arXiv:1903.02493 },
archivePrefix={arXiv},
primaryClass={hep-ph},
url = {https://doi.org/10.48550/arXiv.1903.02493},
}

@article{ref2,
	author = {Li, Zhuang and Wang, Fei},
	title = {Type-II neutrino seesaw mechanism extension of NMSSM from SUSY breaking mechanisms},
	journal = {Eur. Phys. J. C},
	volume = {80},
	year = {2020},
	issn = {1029-8479},
	doi = {10.1140/epjc/s10052-020-8373-0},
	eprint={arXiv:2001.04155},
	archivePrefix={arXiv},
	primaryClass={hep-ph},
	url = {https://doi.org/10.48550/arXiv.2001.04155},
}

@article{ref3,
	title = {Evidence for Oscillation of Atmospheric Neutrinos},
	author = {Fukuda, Y. and Hayakawa, T. and Ichihara, E. and Inoue, K. and Ishihara, K. and Ishino, H. and Itow, Y. and Kajita, T.},
	collaboration = {Super-Kamiokande},
	journal = {Phys. Rev. Lett.},
	volume = {81},
	issue = {8},
	pages = {1562--1567},
	numpages = {0},
	year = {1998},
	month = {Aug},
	publisher = {American Physical Society},
	doi = {10.1103/PhysRevLett.81.1562},
	url = {https://link.aps.org/doi/10.1103/PhysRevLett.81.1562},	
}

@article{ref4,
	title = {Direct Evidence for Neutrino Flavor Transformation from Neutral-Current Interactions in the Sudbury Neutrino Observatory},
	author = {Ahmad, Q. R. and Allen, R. C. and Andersen, T. C. and D.Anglin, J. and Barton, J. C. and Beier},
	collaboration = {SNO},
	journal = {Phys. Rev. Lett.},
	volume = {89},
	issue = {1},
	pages = {011301},
	numpages = {6},
	year = {2002},
	month = {Jun},
	publisher = {American Physical Society},
	doi = {10.1103/PhysRevLett.89.011301},
	eprint={arXiv:nucl-ex/0204008},
	archivePrefix={arXiv},
	primaryClass={hep-ph},
	url = {https://doi.org/10.48550/arXiv.nucl-ex/0204008},
}

@article{ref5,
	title = {First Results from KamLAND: Evidence for Reactor Anti-Neutrino Disappearance},
	author = {Eguchi, K. and Enomoto, S. and Furuno, K. and Goldman, J. and Hanada, H. and Ikeda, H. and Ikeda, K. and Inoue, K.},
	collaboration = {KamLAND},
	journal = {Phys. Rev. Lett.},
	volume = {90},
	issue = {2},
	pages = {021802},
	numpages = {6},
	year = {2003},
	month = {Jan},
	publisher = {American Physical Society},
	doi = {10.1103/PhysRevLett.90.021802},
   eprint={arXiv:hep-ex/0212021},
   archivePrefix={arXiv},
   primaryClass={hep-ph},
   url = {https://doi.org/10.48550/arXiv.hep-ex/0212021},
}

@article{ref6,
	title = {Precision Measurement of Neutrino Oscillation Parameters with KamLAND},
	author = {Abe, S. and Ebihara, T. and Enomoto, S. and Furuno, K. and Gando, Y. and Ichimura, K.},
	collaboration = {KamLAND},
	journal = {Phys. Rev. Lett.},
	volume = {100},
	issue = {22},
	pages = {221803},
	numpages = {5},
	year = {2008},
	month = {Jun},
	publisher = {American Physical Society},
	doi = {10.1103/PhysRevLett.100.221803},
    eprint={arXiv:0801.4589},
    archivePrefix={arXiv},
    primaryClass={hep-ph},
    url = {https://doi.org/10.48550/arXiv.0801.4589},
}

@article{ref7,
	doi = {10.1088/1674-1137/41/7/073103},
	year = {2017},
	month = {jul},
	publisher = {IOP Publishing},
	volume = {41},
	number = {7},
	pages = {073103},
	author = {Xing-Xing Dong and Shu-Min Zhao and Xi-Jie Zhan and Zhong-Jun Yang and Hai-Bin Zhang and Tai-Fu Feng},
	title = { Z $\ensuremath{\rightarrow}$ $\ensuremath{l_i^\pm}$ $\ensuremath{l_j^\pm}$ processes in the BLMSSM},
	journal = {Chinese Physics C},
	eprint={arXiv:1704.02202},
	archivePrefix={arXiv},
	primaryClass={hep-ph},
	url = {https://doi.org/10.48550/arXiv.1704.02202},
}

@article{ref8,
Collaboration = {KATRIN},
title = {Analysis methods for the first KATRIN neutrino-mass measurement},
journal = {Phys. Rev. D},
volume = {104},
issue = {1},
pages = {012005},
numpages = {36},
year = {2021},
month = {Jul},
publisher = {American Physical Society},
doi = {10.1103/PhysRevD.104.012005},
eprint={arXiv:2101.05253 },
archivePrefix={arXiv},
primaryClass={hep-ph},
url = {https://doi.org/10.48550/arXiv.2101.05253},
}

@article{ref9,
Collaboration = {KATRIN},
title = {First direct neutrino-mass measurement with sub-eV sensitivity},
journal = {Nature Physics},
volume = {18},
issue = {2},
pages = {160-166},
year = {2022},
doi = {10.1038/s41567-021-01463-1},
eprint={arXiv:2105.08533},
archivePrefix={arXiv},
primaryClass={hep-ph},
url = {https://doi.org/10.48550/arXiv.2105.08533},
}

@article{refnutrino,
	Collaboration = {KATRIN},
	title = {Direct neutrino-mass measurement based on 259 days of KATRIN data},
	journal = {Science},
	volume = {388},
	number = {6743},
	pages = {180-185},
	year = {2025},
	doi = {10.1126/science.adq9592},
	eprint = {arXiv:2406.13516},
	archivePrefix={arXiv},
	primaryClass={nucl-ex},
	url = {https://doi.org/10.48550/arXiv.2406.13516},
}

@article{ref10,
author = {Hirsch, M. and Joaquim, F. R. and  Vicente, A.},
title = {Constrained SUSY seesaws with a 125 GeV Higgs},
journal = {J. High Energ. Phys.},
volume = {2012},
year = {2012},
doi = {10.1007/JHEP11(2012)105},
eprint={arXiv:1207.6635},
archivePrefix={arXiv},
primaryClass={hep-ph},
url = {https://doi.org/10.48550/arXiv.1207.6635},
}

@article{ref11,
	author = {Esteves, J.N.  and Hirsch, M. and Porod,  W. and Romao,  J.C. and Valle,  J.W.F.  and Villanova del Moral,  A.},
	title = {Flavour violation at the LHC: type-I versus type-II seesaw in minimal supergravity},
	journal = {J. High Energ. Phys.},
	volume = {2009},
	year = {2009},
	doi = {10.1088/1126-6708/2009/05/003},
	eprint={arXiv:0903.1408},
	archivePrefix={arXiv},
	primaryClass={hep-ph},
	url = {https://doi.org/10.48550/arXiv.0903.1408},
}

@article{ref12,
	title = {Baryon- and Lepton-Nonconserving Processes},
	author = {Weinberg, Steven},
	journal = {Phys. Rev. Lett.},
	volume = {43},
	issue = {21},
	pages = {1566-1570},
	numpages = {0},
	year = {1979},
	month = {Nov},
	publisher = {American Physical Society},
	doi = {10.1103/PhysRevLett.43.1566},
	url = {https://link.aps.org/doi/10.1103/PhysRevLett.43.1566},
}

@article{ref13,
	title = {Varieties of baryon and lepton nonconservation},
	author = {Weinberg, Steven},
	journal = {,Phys. Rev. D},
	volume = {22},
	issue = {7},
	pages = {1694-1700},
	numpages = {0},
	year = {1980},
	month = {Oct},
	publisher = {American Physical Society},
	doi = {10.1103/PhysRevD.22.1694},
	url = {https://link.aps.org/doi/10.1103/PhysRevD.22.1694},
}

@article{ref14,
	title = {$\ensuremath{\mu}$$\ensuremath{\rightarrow}$ e $\ensuremath{\gamma}$ at a rate of one out of 109 muon decays?},
	journal = {Physics Letters B},
	volume = {67},
	number = {4},
	pages = {421-428},
	year = {1977},
	issn = {0370-2693},
	doi = {https://doi.org/10.1016/0370-2693(77)90435-X},
	url = {https://www.sciencedirect.com/science/article/pii/037026937790435X},
	author = {Peter Minkowski}
}

@article{ref15,
	author="Yanagida, T.",
	title="Proc. Workshop on Unified Theory and the Baryon Number in the Universe",
	journal="KEK Report No. 79-18",
	year="1979",
	volume="95",
	url="https://cir.nii.ac.jp/crid/1571698599078567552"
}

@inproceedings{ref16,
	title={Talk given at the 1977 Washington meeting of the American Physical Society. See also, M. Gell-Mann, P. Ramond and R. Slansky,“Supergravity”},
	author={Gell-Mann, M},
	booktitle={Proc. of the Supergravity Workshop at Stony Brook, eds P. van Nieuwenhuizen and DZ Freedman (North Holland Publ. Co., Amsterdam, 1979)},
	pages={315}
}

@article{ref17,
	title = {Neutrino Mass and Spontaneous Parity Nonconservation},
	author = {Mohapatra, Rabindra N. and Senjanovi\ifmmode \acute{c}\else \'{c}\fi{}, Goran},
	journal = {Phys. Rev. Lett.},
	volume = {44},
	issue = {14},
	pages = {912-915},
	numpages = {0},
	year = {1980},
	month = {Apr},
	publisher = {American Physical Society},
	doi = {10.1103/PhysRevLett.44.912},
	url = {https://link.aps.org/doi/10.1103/PhysRevLett.44.912}
}

@article{ref18,
	title = {Neutrino masses in SU(2) \ensuremath{\bigotimes} U(1) theories},
	author = {Schechter, J. and Valle, J. W. F.},
	journal = {Phys. Rev. D},
	volume = {22},
	issue = {9},
	pages = {2227-2235},
	numpages = {0},
	year = {1980},
	month = {Nov},
	publisher = {American Physical Society},
	doi = {10.1103/PhysRevD.22.2227},
	url = {https://link.aps.org/doi/10.1103/PhysRevD.22.2227}
}

@article{ref19,
	title = {Neutrino decay and spontaneous violation of lepton number},
	author = {Schechter, J. and Valle, J. W. F.},
	journal = {Phys. Rev. D},
	volume = {25},
	issue = {3},
	pages = {774--783},
	numpages = {0},
	year = {1982},
	month = {Feb},
	publisher = {American Physical Society},
	doi = {10.1103/PhysRevD.25.774},
	url = {https://link.aps.org/doi/10.1103/PhysRevD.25.774}
}

@article{ref20,
	title = {Nonconservation of total lepton number with scalar bosons},
	journal = {Physics Letters B},
	volume = {70},
	number = {4},
	pages = {433-435},
	year = {1977},
	issn = {0370-2693},
	doi = {https://doi.org/10.1016/0370-2693(77)90407-5},
	url = {https://www.sciencedirect.com/science/article/pii/0370269377904075},
	author = {W. Konetschny and W. Kummer}
}

@Inbook{ref21,
	author="Marshak, R. E.
	and Mohapatra, R. N.",
	editor="Perlmutter, Arnold
	and Scott, Linda F.",
	title="Selection Rules for Baryon Number Nonconservation in Gauge Models",
	bookTitle="Recent Developments in High-Energy Physics",
	year="1980",
	publisher="Springer US",
	address="Boston, MA",
	pages="277-287",
	doi="10.1007/978-1-4613-3165-0_18",

}

@article{ref22,
	title = {Neutrino masses, mixings, and oscillations in SU(2)\ifmmode\times\else\texttimes\fi{}U(1) models of electroweak interactions},
	author = {Cheng, T. P. and Li, Ling-Fong},
	journal = {Phys. Rev. D},
	volume = {22},
	issue = {11},
	pages = {2860--2868},
	numpages = {0},
	year = {1980},
	month = {Dec},
	publisher = {American Physical Society},
	doi = {10.1103/PhysRevD.22.2860},
	url = {https://link.aps.org/doi/10.1103/PhysRevD.22.2860}
}

@article{ref23,
	title = {Proton lifetime and fermion masses in an SO(10) model},
	journal = {Nuclear Physics B},
	volume = {181},
	number = {2},
	pages = {287-300},
	year = {1981},
	issn = {0550-3213},
	doi = {https://doi.org/10.1016/0550-3213(81)90354-0},
	url = {https://www.sciencedirect.com/science/article/pii/0550321381903540},
	author = {G. Lazarides and Q. Shafi and C. Wetterich},
}

@article{ref24,
	title = {Neutrino masses and mixings in gauge models with spontaneous parity violation},
	author = {Mohapatra, Rabindra N. and Senjanovi\ifmmode \acute{c}\else \'{c}\fi{}, Goran},
	journal = {Phys. Rev. D},
	volume = {23},
	issue = {1},
	pages = {165--180},
	numpages = {0},
	year = {1981},
	month = {Jan},
	publisher = {American Physical Society},
	doi = {10.1103/PhysRevD.23.165},
	url = {https://link.aps.org/doi/10.1103/PhysRevD.23.165}
}

@article{ref25,
	title = {Proposal for generalised supersymmetry Les Houches Accord for see-saw models and PDG numbering scheme},
	journal = {Computer Physics Communications},
	volume = {184},
	number = {3},
	pages = {698-719},
	year = {2013},
	issn = {0010-4655},
	doi = {https://doi.org/10.1016/j.cpc.2012.11.004},
	author = {L. Basso and A. Belyaev and D. Chowdhury and M. Hirsch and S. Khalil and S. Moretti and B. O’Leary and W. Porod and F. Staub},
	eprint={arXiv:1206.4563},
	archivePrefix={arXiv},
	primaryClass={hep-ph},
	url = {https://doi.org/10.48550/arXiv.1206.4563},
}

@article{ref26,
	title = {See-saw neutrino masses induced by a triplet of leptons},
	author = {R. Foot and H. Lew and X. -G. He and G. C. Joshi },
	journal = {Zeitschrift für Physik C Particles and Fields},
	volume = {44},
	issue = {3},
	pages = {441-444},
	year = {1989},
	doi = {10.1007/BF01415558},
	url = {https://doi.org/10.1007/BF01415558}
}

@article{ref27,
	title = {Pathways to Naturally Small Neutrino Masses},
	author = {Ernest Ma},
	journal = {Phys. Rev. Lett.},
	volume = {81},
	issue = {6},
	pages = {1171--1174},
	numpages = {0},
	year = {1998},
	month = {Aug},
	publisher = {American Physical Society},
	doi = {10.1103/PhysRevLett.81.1171},
	url = {https://link.aps.org/doi/10.1103/PhysRevLett.81.1171}
}

@article{ref28,
author = {Goto, Toru. and Okada, Yasuhiro. and Shindou, Tetsuo. and Tanaka,  Minoru. and Watanabe, Ryoutaro.},
title = {Lepton flavor violation in the supersymmetric seesaw model after the LHC 8 TeV run},
journal = {Phys. Rev. D},
volume = {91},
issue = {3},
pages = {033007},
numpages = {14},
year = {2015},
month = {Feb},
publisher = {American Physical Society},
doi = {10.1103/PhysRevD.91.033007},
eprint={arXiv:1412.2530},
archivePrefix={arXiv},
primaryClass={hep-ph},
url = {https://doi.org/10.48550/arXiv.1412.2530},
}

@article{ref29,
	author = {J. Cao and Z. Xiong and J. M. Yang},
	doi = {10.1140/epjc/s2003-01391-1},
	issn = {1434-6044},
	issue = {2},
	journal = {The European Physical Journal C},
	month = {1},
	pages = {245-252},
	title = {Lepton flavor violating Z-decays in supersymmetric see-saw model},
	volume = {32},
	year = {2004},
	eprint={arXiv:hep-ph/0307126},
	archivePrefix={arXiv},
	primaryClass={hep-ph},
	url = {https://doi.org/10.48550/arXiv.hep-ph/0307126},
}

@article{ref30,
	author = {Darius Jurčiukonis and Luís Lavoura},
	title = {Two-body lepton-flavour-violating decays in a 2HDM with soft family-lepton-number breaking},
	journal = {J. High Energ. Phys.},
	volume = {2022},
	year = {2022},
	doi = {10.1007/JHEP03(2022)106},
	eprint={arXiv:2107.14207},
	archivePrefix={arXiv},
	primaryClass={hep-ph},
	url = {https://doi.org/10.48550/arXiv.2107.14207},
}

@article{ref31,
	author = {Raghavendra Srikanth Hundi},
	title = {Lepton flavor violating Z and Higgs decays in the scotogenic model},
	journal = {Eur. Phys. J. C},
	issue = {6},
	volume = {82,505},
	year = {2022},
	doi = {10.1140/epjc/s10052-022-10453-3},
	eprint={arXiv:2201.03779},
	archivePrefix={arXiv},
	primaryClass={hep-ph},
	url = {https://doi.org/10.48550/arXiv.2201.03779},
}

@article{ref32,
	collaboration = {ATLAS},
	title = {Search for Lepton-Flavor Violation in $Z$-Boson Decays with $\ensuremath{\tau}$ Leptons with the ATLAS Detector},
	journal = {Phys. Rev. Lett.},
	volume = {127},
	issue = {27},
	pages = {271801},
	numpages = {20},
	year = {2021},
	month = {Dec},
	publisher = {American Physical Society},
	doi = {10.1103/PhysRevLett.127.271801},
	eprint={arXiv:2105.12491},
	archivePrefix={arXiv},
	primaryClass={hep-ph},
	url = {https://doi.org/10.48550/arXiv.2105.12491},
}

@article{ref33,
	collaboration = {ATLAS},
	title = {Search for the charged-lepton-flavor-violating decay $Z\ensuremath{\rightarrow}e\ensuremath{\mu}$ in $pp$ collisions at $\sqrt{s}=13\text{ }\text{ }\mathrm{TeV}$ with the ATLAS detector},
	journal = {Phys. Rev. D},
	volume = {108},
	issue = {3},
	pages = {032015},
	numpages = {22},
	year = {2023},
	month = {Aug},
	publisher = {American Physical Society},
	doi = {10.1103/PhysRevD.108.032015},
	eprint={arXiv:2204.10783},
	archivePrefix={arXiv},
	primaryClass={hep-ph},
	url = {https://doi.org/10.48550/arXiv.2204.10783},
}

@article{ref34,
	author = {Calibbi, Lorenzo. and  Marcano, Xabier.  and Roy, Joydeep.},
	title = {Z lepton flavour violation as a probe for new physics at future  $e^+e^-$ colliders},
	journal = {Eur. Phys. J. C},
	year = {2021},
	volume = {81},
	pages = {1054},
	doi = {10.1140/epjc/s10052-021-09777-3},
	eprint={arXiv:2107.10273},
	archivePrefix={arXiv},
	primaryClass={hep-ph},
	url = {https://doi.org/10.48550/arXiv.2107.10273},
}

@article{ref35,
	author = {G. Bernardi and E. Brost and D. Denisov and G. Landsberg and M. Aleksa and D. d'Enterria and P. Janot and M. L. Mangano and M. Selvaggi and F. Zimmermann and J. Alcaraz Maestre and C. Grojean and R. M. Harris and A. Pich and M. Vos and S. Heinemeyer and P. Giacomelli and P. Azzi and F. Bedeschi and M. Klute and A. Blondel and C. Paus and F. Simon and M. Dam and E. Barberis and L. Skinnari and T. Raubenheimer and S. Antusch and W. Altmannshofer and L. -T. Wang and J. de Blas and S. Eno and Yihui Lai and S. Willocq and J. Qian and J. Zhu and R. Novotny and S. Seidel and M. D. Hildreth and E. J. Thomson and R. Demina and J. Gluza and G. Isidori and R. Gonzalez Suarez},
	title = {The Future Circular Collider: a Summary for the US 2021 Snowmass Process},
	year = {2022},
	eprint={arXiv:2203.06520},
	archivePrefix={arXiv},
	primaryClass={hep-ph},
	url = {https://doi.org/10.48550/arXiv.2203.06520},
}

@article{ref36,
	title = {Lepton flavor violation in Z and lepton decays in supersymmetric models},
	author = {Illana, J. I. and Masip, M.},
	journal = {Phys. Rev. D},
	volume = {67},
	issue = {3},
	pages = {035004},
	numpages = {12},
	year = {2003},
	month = {Feb},
	publisher = {American Physical Society},
	doi = {10.1103/PhysRevD.67.035004},
	eprint={arXiv:hep-ph/0207328},
	archivePrefix={arXiv},
	primaryClass={hep-ph},
	url = {https://doi.org/10.48550/arXiv.hep-ph/0207328},
	}

@article{ref37,
	author = {Hirsch, M. and Porod,  W. and Weiss, Ch. and Staub, F.},
	title = {Supersymmetric type-III seesaw: lepton flavour violation and LHC phenomenology},
	volume = {87},
	journal = {Phys. Rev. D},
	year = {2013},
	pages = {013010},
	doi = {10.1103/PhysRevD.87.013010},
	eprint={arXiv:1211.0289 },
	archivePrefix={arXiv},
	primaryClass={hep-ph},
	url = {https://doi.org/10.48550/arXiv.1211.0289},
}

@article{ref38,
	author = {Hirsch, M. and Kaneko,  S. and Porod,  W.},
	title = {Supersymmetric seesaw type II: CERN LHC and lepton flavour violating phenomenology},
	volume = {78},
	journal = {Phys. Rev. D},
	year = {2008},
	pages = {093004},
	doi = {10.1103/PhysRevD.78.093004},
	eprint={arXiv:0806.3361},
	archivePrefix={arXiv},
	primaryClass={hep-ph},
	url = {https://doi.org/10.48550/arXiv.0806.3361},
}

@article{ref39,
	author = {Borzumati, Francesca. and Yamashita, Toshifumi.},
	title = {Minimal supersymmetric SU(5) model with nonrenormalizable operators: Seesaw mechanism and violation of flavour and CP},
	journal = {Progress of Theoretical Physics},
	volume = {124},
	number = {5},
	pages = {761-868},
	year = {2010},
	month = {11},
	issn = {0033-068X},
	doi = {10.1143/PTP.124.761},
	eprint={arXiv:0903.2793},
	archivePrefix={arXiv},
	primaryClass={hep-ph},
	url = {https://doi.org/10.1143/PTP.124.761}
}

@article{ref40,
	author = {A. Abada and A. J. R. Figueiredo and J. C. Romão and A. M. Teixeira },
	title = {Probing the supersymmetric type III seesaw: LFV at low-energies and at the LHC},
	journal = {J. High Energ. Phys.},
	volume = {2011},
	year = {2011},
	doi = {10.1007/JHEP08(2011)099},
	eprint={arXiv:1104.3962},
archivePrefix={arXiv},
primaryClass={hep-ph},
url = {https://doi.org/10.48550/arXiv.1104.3962},
}

@article{ref41,
	title = {Supersymmetric type-III seesaw: lepton flavour violating decays and dark matter},
	author = {Esteves, J. N. and Romao, J. C. and Hirsch, M. and Staub, F. and Porod, W.},
	journal = {Phys. Rev. D},
	volume = {83},
	issue = {1},
	pages = {013003},
	numpages = {21},
	year = {2011},
	month = {Jan},
	publisher = {American Physical Society},
	doi = {10.1103/PhysRevD.83.013003},
eprint={arXiv:1010.6000},
archivePrefix={arXiv},
primaryClass={hep-ph},
url = {https://doi.org/10.48550/arXiv.1010.6000},
}

@article{ref42,
	author ={A. Vicente},
	title = {Lepton flavor violation beyond the MSSM},
	journal = {Advances in High Energy Physics},
	volume = {2015},
	pages = {22},
	year = {2015},
	doi = {10.1155/2015/686572},
	eprint={arXiv:1503.08622},
	archivePrefix={arXiv},
	primaryClass={hep-ph},
	url = {https://doi.org/10.48550/arXiv.1503.08622},
}

@article{ref43,
	author ={Goodsell, Mark D. and Nickel, Kilian. and Staub, F.},
	title = {Two-Loop Higgs mass calculations in supersymmetric models beyond the MSSM with SARAH and SPheno},
	journal = {Eur. Phys. J. C},
	volume = {75},
	pages = {}, 
	year = {2015},
	doi = {10.1140/epjc/s10052-014-3247-y},
	eprint={arXiv:1411.0675},
	archivePrefix={arXiv},
	primaryClass={hep-ph},
	url = {https://doi.org/10.48550/arXiv.1411.0675},
}

@article{ref44,
author ={Bernigaud, Jordan. and Forster, Adam K. and Herrmann, Björn. and King, Stephen F. and Porod, Werner. and Rowley, Samuel J.},
title = {Data-driven analysis of a SUSY GUT of flavour},
journal = {J. High Energ. Phys.},
volume = {2022},
pages = {},
year = {2022},
doi = {10.1007/JHEP05(2022)156},
eprint={arXiv:2111.10199},
archivePrefix={arXiv},
primaryClass={hep-ph},
url = {https://doi.org/10.48550/arXiv.2111.10199},
}

@article{ref45,
author ={Porod, W. and Staub, F. and Vicente,  A.},
title = {A Flavor Kit for BSM models},
journal = {Eur. Phys. J. C},
volume = {74},
pages = {},
year = {2014},
doi = {10.1140/epjc/s10052-014-2992-2},
eprint={arXiv:1405.1434},
archivePrefix={arXiv},
primaryClass={hep-ph},
url = {https://doi.org/10.48550/arXiv.1405.1434},
}

@article{ref46,
	author = {D'Ambrosio, Giancarlo. and Hambye, Thomas. and Hektor, Andi. and Raidal, Martti. and Rossi, Anna.},
	title = {Leptogenesis in the minimal supersymmetric triplet seesaw model},
	journal = {Physics Letters B},
	volume = {604},
	number = {3},
	pages = {199-206},
	year = {2004},
	issn = {0370-2693},
	doi = {https://doi.org/10.1016/j.physletb.2004.10.056},
	eprint={arXiv:hep-ph/0407312},
	archivePrefix={arXiv},
	primaryClass={hep-ph},
	url = {https://doi.org/10.48550/arXiv.hep-ph/0407312},
}

@article{ref47,
	author = {Joaquim, F.R. and Rossi, Anna.},
	title = {Phenomenology of the triplet seesaw mechanism with Gauge and Yukawa mediation of SUSY breaking},
	journal = {Nuclear Physics B},
	volume = {765},
	number = {1},
	pages = {71-117},
	year = {2007},
	issn = {0550-3213},
	doi = {https://doi.org/10.1016/j.nuclphysb.2006.11.030},
	eprint={arXiv:hep-ph/0607298},
	archivePrefix={arXiv},
	primaryClass={hep-ph},
	url = {https://doi.org/10.48550/arXiv.hep-ph/0607298},
}

@article{ref48,
	author = {Rossi, Anna.},
	title = {Supersymmetric Seesaw without Singlet Neutrinos: Neutrino Masses and Lepton-Flavour Violation},
	journal = {Phys. Rev. D},
	volume = {66},
	issue = {7},
	pages = {075003},
	numpages = {17},
	year = {2002},
	month = {Oct},
	publisher = {American Physical Society},
	doi = {10.1103/PhysRevD.66.075003},
	eprint={arXiv:hep-ph/0207006},
	archivePrefix={arXiv},
	primaryClass={hep-ph},
	url = {https://doi.org/10.48550/arXiv.hep-ph/0207006},
}

@article{ref49,
	author = {CS\'{A}KI, CSABA},
	title = {The Minimal Supersymmetric Standard Model (MSSM)},
	journal = {Modern Physics Letters A},
	volume = {11},
	number = {08},
	pages = {599-613},
	year = {1996},
	doi = {10.1142/S021773239600062X},
	eprint={arXiv:hep-ph/9606414},
	archivePrefix={arXiv},
	primaryClass={hep-ph},
	url = {https://doi.org/10.48550/arXiv.hep-ph/9606414},
}

@article{ref50,
	author = {JinMin Yang},
	doi = {10.1007/s11433-010-4146-3},
	issn = {1674-7348},
	issue = {11},
	journal = {Science China Physics, Mechanics and Astronomy},
	month = {11},
	pages = {1949-1952},
	title = {Lepton flavor violating Z-boson decays at GigaZ as a probe of supersymmetry},
	volume = {53},
	year = {2010},
	eprint={arXiv:1006.2594},
	archivePrefix={arXiv},
	primaryClass={hep-ph},
	url = {https://doi.org/10.48550/arXiv.1006.2594},
}

@article{ref51,
	author = {De Romeri, V. and Herrero,  M.J. and Marcano,  X. and Scarcella, F.},
	title = {Lepton flavor violating Z decays: A promising window to low scale seesaw neutrinos},
	journal = {Phys. Rev. D},
	volume = {95},
	issue = {7},
	pages = {075028},
	numpages = {22},
	year = {2017},
	month = {Apr},
	publisher = {American Physical Society},
	doi = {10.1103/PhysRevD.95.075028},
	eprint={arXiv:1607.05257},
	archivePrefix={arXiv},
	primaryClass={hep-ph},
	url = {https://doi.org/10.48550/arXiv.1607.05257},
}

@book{ref52,
	author={Marcano, Xabier.},
	title = {Lepton Flavor Violation from Low Scale Seesaw Neutrinos with Masses Reachable at the LHC},
	year={2018},
	school ={The Autonomous University of Madrid, Spain},
	pages={162} , 
	Publisher={Springer Cham},  
	doi = {10.1007/978-3-319-94604-7},
	url = {https://doi.org/10.1007/978-3-319-94604-7}
}

@article{ref53,
	doi = {10.1088/1674-1137/43/4/043101},
	year = {2019},
	month = {apr},
	publisher = {Chinese Physical Society and the Institute of High Energy Physics of the Chinese Academy of Sciences and the Institute of Modern Physics of the Chinese Academy of Sciences and IOP Publishing Ltd},
	volume = {43},
	number = {4},
	pages = {043101},
	author = {Ke-Sheng Sun and Jian-Bin Chen and Xiu-Yi Yang and Sheng-Kai Cui},
	title = {The LFV decays of Z boson in Minimal R-symmetric Supersymmetric Standard Model},
	journal = {Chinese Physics C},
	eprint={arXiv:1901.03800},
	archivePrefix={arXiv},
	primaryClass={hep-ph},
	url = {https://doi.org/10.48550/arXiv.1901.03800},
}

@article{ref54,
	title = {Lepton flavor violation in supersymmetric SO(10) grand unified models},
	author = {Bi, Xiao-Jun and Dei, Yuan-Ben and Qi, Xiao-Yuan},
	journal = {Phys. Rev. D},
	volume = {63},
	issue = {9},
	pages = {096008},
	numpages = {13},
	year = {2001},
	month = {Apr},
	publisher = {American Physical Society},
	doi = {10.1103/PhysRevD.63.096008},
	eprint={arXiv:hep-ph/0010270},
	archivePrefix={arXiv},
	primaryClass={hep-ph},
	url = {https://doi.org/10.48550/arXiv.hep-ph/0010270}
}

@article{ref55,
	title = {Review of Particle Physics},
	author = {Navas, S. and Amsler, C. and Gutsche, T. and Hanhart, C. and Hern\'andez-Rey, J. J. and Louren\ifmmode \mbox{\c{c}}\else \c{c}\fi{}o, C. and Masoni, A. and Mikhasenko, M. and Mitchell, R. E. and Patrignani, C. and Schwanda, C. and Spanier, S. and Venanzoni, G. and Yuan, C. Z. and Agashe, K. and Aielli, G. and Allanach, B. C. and Alvarez-Mu\~niz, J. and Antonelli, M. and Aschenauer, E. C. and Asner, D. M. and Assamagan, K. and Baer, H. and Banerjee, Sw. and Barnett, R. M. and Baudis, L. and Bauer, C. W. and Beatty, J. J. and Beringer, J. and Bettini, A. and Biebel, O. and Black, K. M. and Blucher, E. and Bonventre, R. and Briere, R. A. and Buckley, A. and Burkert, V. D. and Bychkov, M. A. and Cahn, R. N. and Cao, Z. and Carena, M. and Casarosa, G. and Ceccucci, A. and Cerri, A. and Chivukula, R. S. and Cowan, G. and Cranmer, K. and Crede, V. and Cremonesi, O. and D'Ambrosio, G. and Damour, T. and de Florian, D. and de Gouv\^ea, A. and DeGrand, T. and Demers, S. and Demiragli, Z. and Dobrescu, B. A. and D'Onofrio, M. and Doser, M. and Dreiner, H. K. and Eerola, P. and Egede, U. and Eidelman, S. and El-Khadra, A. X. and Ellis, J. and Eno, S. C. and Erler, J. and Ezhela, V. V. and Fava, A. and Fetscher, W. and Fields, B. D. and Freitas, A. and Gallagher, H. and Gershon, T. and Gershtein, Y. and Gherghetta, T. and Gonzalez-Garcia, M. C. and Goodman, M. and Grab, C. and Gritsan, A. V. and Grojean, C. and Groom, D. E. and Gr\"unewald, M. and Gurtu, A. and Haber, H. E. and Hamel, M. and Hashimoto, S. and Hayato, Y. and Hebecker, A. and Heinemeyer, S. and Hikasa, K. and Hisano, J. and H\"ocker, A. and Holder, J. and Hsu, L. and Huston, J. and Hyodo, T. and Ianni, Al. and Kado, M. and Karliner, M. and Katz, U. F. and Kenzie, M. and Khoze, V. A. and Klein, S. R. and Krauss, F. and Kreps, M. and Kri\ifmmode \check{z}\else \v{z}\fi{}an, P. and Krusche, B. and Kwon, Y. and Lahav, O. and Lellouch, L. P. and Lesgourgues, J. and Liddle, A. R. and Ligeti, Z. and Lin, C.-J. and Lippmann, C. and Liss, T. M. and Lister, A. and Littenberg, L. and Lugovsky, K. S. and Lugovsky, S. B. and Lusiani, A. and Makida, Y. and Maltoni, F. and Manohar, A. V. and Marciano, W. J. and Matthews, J. and Mei\ss{}ner, U.-G. and Melzer-Pellmann, I.-A. and Mertsch, P. and Miller, D. J. and Milstead, D. and M\"onig, K. and Molaro, P. and Moortgat, F. and Moskovic, M. and Nagata, N. and Nakamura, K. and Narain, M. and Nason, P. and Nelles, A. and Neubert, M. and Nir, Y. and O'Connell, H. B. and O'Hare, C. A. J. and Olive, K. A. and Peacock, J. A. and Pianori, E. and Pich, A. and Piepke, A. and Pietropaolo, F. and Pomarol, A. and Pordes, S. and Profumo, S. and Quadt, A. and Rabbertz, K. and Rademacker, J. and Raffelt, G. and Ramsey-Musolf, M. and Richardson, P. and Ringwald, A. and Robinson, D. J. and Roesler, S. and Rolli, S. and Romaniouk, A. and Rosenberg, L. J and Rosner, J. L. and Rybka, G. and Ryskin, M. G. and Ryutin, R. A. and Safdi, B. and Sakai, Y. and Sarkar, S. and Sauli, F. and Schneider, O. and Sch\"onert, S. and Scholberg, K. and Schwartz, A. J. and Schwiening, J. and Scott, D. and Sefkow, F. and Seljak, U. and Sharma, V. and Sharpe, S. R. and Shiltsev, V. and Signorelli, G. and Silari, M. and Simon, F. and Sj\"ostrand, T. and Skands, P. and Skwarnicki, T. and Smoot, G. F. and Soffer, A. and Sozzi, M. S. and Spiering, C. and Stahl, A. and Sumino, Y. and Takahashi, F. and Tanabashi, M. and Tanaka, J. and Ta\ifmmode \check{s}\else \v{s}\fi{}evsk\'y, M. and Terao, K. and Terashi, K. and Terning, J. and Thoma, U. and Thorne, R. S. and Tiator, L. and Titov, M. and Tovey, D. R. and Trabelsi, K. and Urquijo, P. and Valencia, G. and Van de Water, R. and Varelas, N. and Verde, L. and Vivarelli, I. and Vogel, P. and Vogelsang, W. and Vorobyev, V. and Wakely, S. P. and Walkowiak, W. and Walter, C. W. and Wands, D. and Weinberg, D. H. and Weinberg, E. J. and Wermes, N. and White, M. and Wiencke, L. R. and Willocq, S. and Woody, C. L. and Workman, R. L. and Yao, W.-M. and Yokoyama, M. and Yoshida, R. and Zanderighi, G. and Zeller, G. P. and Zhu, R.-Y. and Zhu, S.-L. and Zimmermann, F. and Zyla, P. A. and Anderson, J. and Kramer, M. and Schaffner, P. and Zheng, W.},
	journal = {Phys. Rev. D},
	volume = {110},
	issue = {3},
	pages = {030001},
	numpages = {5},
	year = {2024},
	month = {Aug},
	publisher = {American Physical Society},
	doi = {10.1103/PhysRevD.110.030001},
	url = {https://link.aps.org/doi/10.1103/PhysRevD.110.030001},
}

@article{ref56,
	title = {One-loop corrections for $e^+e^-$ annihilation into $\ensuremath{\mu^+}$$\ensuremath{\mu^-}$ in the Weinberg model},
	journal = {Nuclear Physics B},
	volume = {160},
	number = {1},
	pages = {151-207},
	year = {1979},
	issn = {0550-3213},
	doi = {https://doi.org/10.1016/0550-3213(79)90234-7},
	url = {https://www.sciencedirect.com/science/article/pii/0550321379902347},
	author = {G. Passarino and M. Veltman},
}

@article{ref57,
	title = {Automatic Loop Calculations with FeynArts, FormCalc, and LoopTools},
	author = {Thomas Hahn},
	journal = {Nuclear Physics B - Proceedings Supplements},
	volume = {89},
	number = {1},
	pages = {231-236},
	year = {2000},
	note = {Loops and Legs in Quantum Field Theory},
	issn = {0920-5632},
	doi = {https://doi.org/10.1016/S0920-5632(00)00848-3},
   eprint={arXiv:hep-ph/0005029},
   archivePrefix={arXiv},
   primaryClass={hep-ph},
   url = {https://doi.org/10.48550/arXiv.0005029},
}

@article{ref58,
	title = {Package-X: A Mathematica package for the analytic calculation of one-loop integrals},
	journal = {Computer Physics Communications},
	volume = {197},
	pages = {276-290},
	year = {2015},
	issn = {0010-4655},
	doi = {https://doi.org/10.1016/j.cpc.2015.08.017},
	author = {Hiren H. Patel},
	eprint={arXiv:1503.01469},
	archivePrefix={arXiv},
	primaryClass={hep-ph},
	url = {https://doi.org/10.48550/arXiv.1503.01469},
}

@article{ref59,
	author={Staub, F.},
	title = {From Superpotential to Model Files for FeynArts and CalcHep/CompHep},
	journal={Computer.Physics.Commun.},
	volume={181},
	pages={1077-1086},
	year={2010},
	doi = {10.1016/j.cpc.2010.01.011},
	eprint={arXiv:0909.2863},
	archivePrefix={arXiv},
	primaryClass={hep-ph},
	url = {https://doi.org/10.48550/arXiv.0909.2863},
}

@article{ref60,
	author ={Staub, F.},
	title = {SARAH 4: A tool for (not only SUSY) model builders},
	journal = {Computer.Physics.Commun.},
	volume = {185},
	pages = {1773-1790},
	year = {2014},
	doi = {10.1016/j.cpc.2014.02.018},
	eprint={arXiv:1309.7223},
	archivePrefix={arXiv},
	primaryClass={hep-ph},
	url = {https://doi.org/10.48550/arXiv.1309.7223},
}

@article{ref61,
	author = {},
	Collaboration = {Belle},
	title = {Search for lepton-flavor-violating tau-lepton decays to $\ensuremath{\ell}$$\ensuremath{\gamma}$ at Belle},
	journal = {J. High Energ. Phys.},
	volume = {2021},
	year = {2021},
	doi = {10.1007/JHEP10(2021)019},
eprint={arXiv:2103.12994},
archivePrefix={arXiv},
primaryClass={hep-ph},
url = {https://doi.org/10.48550/arXiv.2103.12994},
}

@article{ref62,
	author = {},
	Collaboration = {MEG},
	title = {Search for the Lepton Flavour Violating Decay $\ensuremath{\mu^+}$ $\ensuremath{\rightarrow}$ $\ensuremath{e^+}$ $\ensuremath{\gamma}$ with the Full Dataset of the MEG Experiment},
	journal = {, Eur. Phys. J. C},
	volume = {76},
	year = {2016},
	issn = {434},
	doi = {10.1140/epjc/s10052-016-4271-x},
	eprint={arXiv:1605.05081},
	archivePrefix={arXiv},
	primaryClass={hep-ph},
	url = {https://doi.org/10.48550/arXiv.1605.05081},
	}

@article{ref63,
	title = {Lepton-Flavor Violation via Right-Handed Neutrino Yukawa Couplings in Supersymmetric Standard Model},
	author = {Hisano, J. and Moroi, T. and Tobe, K. and Yamaguchi, M.},
	journal = {Phys. Rev. D},
	volume = {53},
	issue = {5},
	pages = {2442--2459},
	numpages = {0},
	year = {1996},
	month = {Mar},
	publisher = {American Physical Society},
	doi = {10.1103/PhysRevD.53.2442},
	eprint={arXiv:hep-ph/9510309},
	archivePrefix={arXiv},
	primaryClass={hep-ph},
	url = {https://doi.org/10.48550/arXiv.9510309},
	}

@Article{ref64,
	AUTHOR = {Israr, Sajid and Gómez, Mario E. and Rehman, Muhammad},
	TITLE = {Nonholomorphic Higgsino Mass Term Effects on Muon $g-2$ and Dark Matter Relic Density in Flavor Symmetry Based Minimal Supersymmetric Standard Model},
	JOURNAL = {Particles},
	VOLUME = {8},
	YEAR = {2025},
	NUMBER = {1},
	ARTICLE-NUMBER = {30},
	URL = {https://www.mdpi.com/2571-712X/8/1/30},
	ISSN = {2571-712X},
	DOI = {10.3390/particles8010030}
}

\end{document}